\documentclass[5p]{elsarticle}

\usepackage{graphicx}
\usepackage{ulem}
\usepackage{amsmath}
\usepackage{comment}
\usepackage[hidelinks]{hyperref}

\journal{Cryogenics}
\begin{document}


\begin{frontmatter}

\title{The Simons Observatory: Design and Measured Performance of a \\ Carbon Fiber Strut for a Cryogenic Truss}

\author[princeton]{Kevin D.\ Crowley}
\author[uva]{Peter Dow}
\author[uva]{Jordan E.\ Shroyer} 
\author[nist]{John C.\ Groh}
\author[cu,nist]{Bradley Dober}
\author[ucsd]{Jacob Spisak}
\author[ucsd]{Nicholas Galitzki}
\author[penn]{Tanay Bhandarkar} 
\author[penn]{Mark J.\ Devlin} 
\author[penn]{Simon Dicker}
\author[chicago1]{Patricio A.\ Gallardo}
\author[chicago2]{Kathleen Harrington}
\author[uva]{Bradley R.\ Johnson\corref{corauth}} \cortext[corauth]{Corresponding author} \ead{bradley.johnson@virginia.edu} 
\author[ucsd]{Delwin Johnson}
\author[penn]{Anna M.\ Kofman} 
\author[lbnl,tokyo1,berkeley1,tokyo2]{Akito Kusaka}
\author[berkeley2]{Adrian Lee} 
\author[penn]{Michele Limon} 
\author[penn]{Jeffrey Iuliano} 
\author[milano]{Federico Nati} 
\author[penn]{John Orlowski-Scherer} 
\author[princeton]{Lyman Page} 
\author[ucsd]{Michael Randall}
\author[ucsd]{Grant Teply}
\author[ucsd]{Tran Tsan} 
\author[gsfc]{Edward J.\ Wollack} 
\author[mit]{Zhilei Xu} 
\author[penn]{Ningfeng Zhu} 

\address[princeton]{Joseph Henry Laboratories of Physics, Jadwin Hall, Princeton University, Princeton, NJ 08544, USA} 
\address[uva]{University of Virginia, Department of Astronomy, Charlottesville, VA 22904, USA}
\address[nist]{National Institute of Standards and Technology, Boulder, CO 80305, USA} 
\address[cu]{University of Colorado Boulder, Department of Physics, Boulder, CO 80309, USA}
\address[ucsd]{University of California, San Diego, Department of Physics, San Diego, CA 92093, USA}
\address[penn]{Department of Physics and Astronomy, University of Pennsylvania, 209 South 33rd St., Philadelphia, PA 19104, USA}
\address[chicago1]{Kavli Institute for Cosmological Physics, University of Chicago, Chicago, IL 60637, USA}
\address[chicago2]{Department of Astronomy and Astrophysics, University of Chicago, Chicago, IL 60637, USA}
\address[lbnl]{Physics Division, Lawrence Berkeley National Laboratory, Berkeley, CA 94720, USA}
\address[tokyo1]{Department of Physics, The University of Tokyo, Tokyo 113-0033, Japan}
\address[berkeley1]{Kavli Institute for the Physics and Mathematics of the Universe (WPI), Berkeley Satellite, University of California, Berkeley 94720, USA}
\address[tokyo2]{Research Center for the Early Universe, School of Science, The University of Tokyo, Tokyo 113-0033, Japan}
\address[berkeley2]{Physics Department, University of California, Berkeley, CA 94720, USA}
\address[milano]{Department of Physics, University of Milano-Bicocca, Piazza della Scienza 3, Milan (MI), 20126, Italy}
\address[gsfc]{NASA Goddard Space Flight Center, Greenbelt, MD 20771, USA}
\address[mit]{MIT Kavli Institute, Massachusetts Institute of Technology, 77 Massachusetts Avenue, Cambridge, MA 02139, USA}

\begin{abstract}
We present the design and measured performance of a new carbon fiber strut design that is used in a cryogenically cooled truss for the Simons Observatory Small Aperture Telescope (SAT).
The truss consists of two aluminum 6061 rings separated by 24 struts.
Each strut consists of a central carbon fiber tube fitted with two aluminum end caps.
We tested the performance of the strut and truss by (i) cryogenically cycling and destructively pull-testing strut samples, (ii) non-destructively pull-testing the final truss, and (iii) measuring the thermal conductivity of the carbon fiber tubes.
We found that the strut strength is limited by the mounting fasteners and the strut end caps, not the epoxy adhesive or the carbon fiber tube.
This result is consistent with our numerical predictions.
Our thermal measurements suggest that the conductive heat load through the struts (from 4~K to 1~K) will be less than 1~mW.
This strut design may be a promising candidate for use in other cryogenic support structures.
\end{abstract}


\end{frontmatter}




\section{Introduction} 
\label{sec:introduction}

Instruments for measuring the cosmic microwave background (CMB) have seen steady improvement in sensitivity in recent decades~\cite{Staggs_2018}. 
These improvements have primarily been achieved by using larger detector arrays with more array elements and optical systems that are cooled to cryogenic temperatures to minimize unwanted thermal emission from the instrument~\cite{mccarrick_2021, sobrin_2021, moncelsi_2020, dahal_2020, kiuchi_2020, ade_2019, galitzki_2018, datta_2016}.
These two instrument features have created a need for cryomechanical systems that are capable of supporting the steadily increasing size and mass of these instrument components.

A common approach is to support the instrument components inside the cryostat with a truss system composed of struts~\cite{sobrin_2021, moncelsi_2020, kiuchi_2020, galitzki_2018, zhu_2021, groh_2021, Howe_2018, gudmundsson_2015, schwan_2011}.
If the struts are properly designed, this approach delivers both the mechanical strength needed to support and align the instrument components, and the thermal isolation needed to cool the various temperature stages of the instrument.
Composites including carbon fiber reinforced polymers (CFRP) have been shown to work well for this kind of application~\cite{zsombor_2020}.
In this paper we present both the design and the measured performance of a new strut that is used in a cryogenically cooled truss for the Simons Observatory (SO) Small Aperture Telescope (SAT)~\cite{kiuchi_2020, galitzki_2018}.

We motivate our design by describing one of the five cryogenic trusses in the SAT in Section~\ref{sec:truss_design}.
The mechanical and the thermal requirements for the struts were set using this truss, and these requirements are given in Section~\ref{sec:performance_requirements}.
We then describe the strut design in Section~\ref{sec:strut_design} and the expected performance in Section~\ref{sec:calculated_strength}.
To ascertain the ultimate strength of the struts, we made strut samples and pull tested them to failure.
These measurements are described in Section~\ref{sec:strut_testing}.
We then assembled a truss using the struts and tested the full assembly in a variety of ways.
These measurements are described in Section~\ref{sec:truss_testing}.
We also measured the thermal conductivity of the carbon fiber tubes and present results in Section~\ref{sec:thermal_testing}.
Our study validates the design philosophy, assembly procedure, and chosen materials.


\section{Methods}
\label{sec:methods}


\begin{figure}[t]
\centering
\includegraphics[width=\columnwidth]{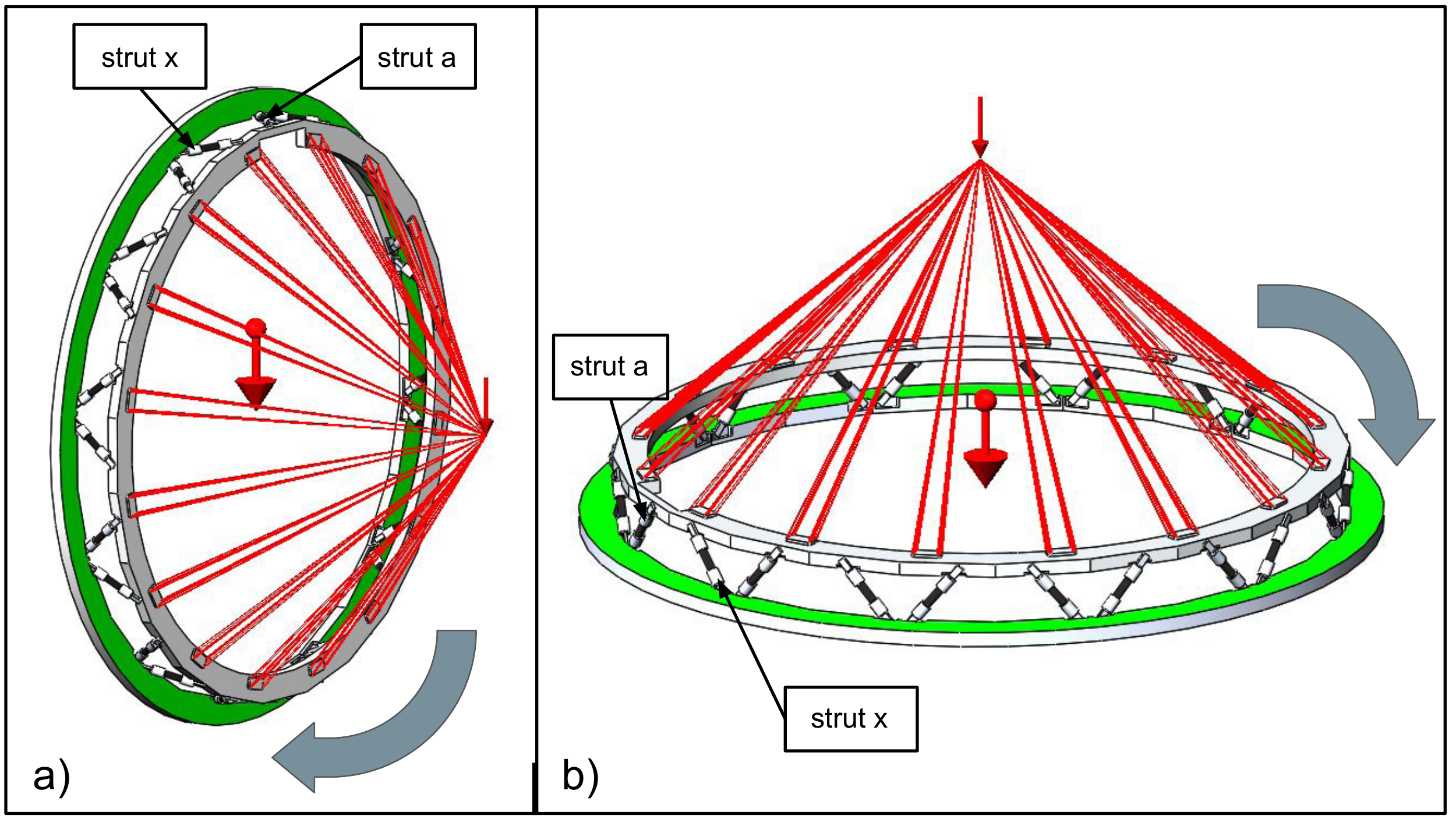}
\caption{
Simulation setups for configurations~\#3 (panel $a$) and \#5 (panel $b$) as described in the text.
The green surface of the 4~K ring is fixed in the simulation.
The total mass of the instrument components mounted to the 1~K truss ring is 215~kg, and the associated center of mass is displaced 31.2~cm from the mounting surface of the 1~K ring.
The direction of the load is indicated by the small red arrow, and the rigid fixtures are indicated by the red lines connecting the load to the 1~K ring of the truss.
The direction of gravity is indicated by the large red arrow in the center of the truss.
Struts $a$ and $x$ are called out in both panels, and the direction of increasing strut letter ($a$ to $x$) is indicated by the thick grey arrow wrapping around the truss.
Truss dimensions are given in Table~\ref{table:trussRingDimensions}.
}
\label{fig:TrussSimOrientation}
\end{figure}


\begin{table}

\footnotesize
\centering
\renewcommand\arraystretch{1.3}

\begin{tabular}{cccc}
\hline
Stage & OD [\,cm\,] & ID [\,cm\,] & Thickness [\,cm\,] \\
\hline
4~K & 82.0 & 74.3 & 1.5 \\
1~K & 74.5 & 68.0 & 1.4 \\
\hline
\end{tabular}
\caption{4~K and 1~K truss ring dimensions.}
\label{table:trussRingDimensions}
\end{table}


\subsection{Truss Description}
\label{sec:truss_design}

The truss consists of two aluminum 6061 rings separated by 24 carbon fiber struts. 
The larger-diameter ring is thermally connected to the second stage of a Cryomech\footnote{Cryomech Inc, 6682 Moore Rd., Syracuse, NY 13211 USA} PT420 pulse tube cooler (PTC), henceforth referred to as the 4~K stage of the instrument, and the smaller diameter ring is thermally connected to the still of a BlueFors\footnote{Bluefors Oy, Arinatie 10, 00370 Helsinki, Finland} SD400 dilution refrigerator (DR), henceforth referred to as the 1~K stage.
The dimensions of the truss rings are given in Table~\ref{table:trussRingDimensions}.
The struts are mounted 45~deg from the planes of the truss rings, which are separated by 48.17~mm. 


\subsection{Performance Requirements}
\label{sec:performance_requirements}

The truss must be mechanically strong enough to support all of the instrument components mounted to the 1~K stage of the instrument.
This includes both the telescope optics, composed of silicon lenses~\cite{kiuchi_2020, galitzki_2018, datta_2013}, and the detector system, composed of seven universal focal-plane modules~\cite{mccarrick_2021}.
The total mass of these components is 215~kg.
Additionally, the thermal load through the struts onto the 1~K stage must be compatible with the overall thermal budget of the instrument.
The DR has a measured still cooling power of 30~mW at 1.2~K, so any thermal load on the 1~K stage through the struts must be much less than 30~mW; we targeted less than 5\% of the total cooling power as a requirement for the parasitic load.
Regarding the mechanical strength requirement, we used the finite element analysis (FEA) extension of SolidWorks\footnote{Dassault Systèmes SolidWorks Corporation, 175 Wyman Street, Waltham, MA 02451, USA} to compute the expected loads on each strut in five critical orientations.
The maximum loads from these simulations provided the strength requirements for the struts.

The five truss configurations correspond to the telescope elevation angles $-$90, $-$45, 0, 45, and 90~deg, and they are referred to as configurations~\#1 to \#5, respectively.
An elevation angle of 90~deg points the telescope up at the zenith, while an elevation angle of $-$90~deg points the telescope down at the ground.
These five orientations form a basis upon which all other maintenance and observing orientations of the instrument can be decomposed. 
We also expect that the maximum strut loads will occur at these orientations because they are points at which the projection of the gravity vector onto the central axes of either the truss or the struts is maximized, minimized, or zero.

Configuration~\#1 (el = $-$90~deg, telescope pointing down) puts the struts in tension with the 1~K ring being pulled away from the fixed 4~K ring.
Configurations~\#2 to \#4 (el = $-$45, 0, and 45~deg, respectively) put the struts in a mixture of tension and compression.
Configuration~\#5 (el = 90~deg, telescope pointing up) puts the struts in compression with the 1~K ring being pushed down onto the fixed 4~K ring.
We expect to perform routine observations between 20 and 70~deg, so configuration~\#4 corresponds to an observing configuration.
The other configurations correspond to scenarios not related to observation, including telescope maintenance and transport.

Configuration~\#5 is special because the instrument is transported across the observatory in this orientation.
Therefore, we imposed more stringent requirements for this configuration.
Here we assume the magnitude of the load is 10 times the load under 1$g$ to simulate the acceleration associated with an unexpected shock to the instrument during transport.
This is intended as an extreme case given several protocols are in place to ensure shock loads do not exceed 3$g$ in reality.


\begin{table}
\footnotesize
\centering
\renewcommand\arraystretch{1.3}
\begin{tabular}{ccc}
\hline
Force type & Force [\,N\,] & Configuration    \\    
\hline
Tension & 372. & 3 \\
Compression & 1450 & 5 \\
Shear & 35.9 & 5 \\
\hline
\end{tabular}
\caption{
The maximum tension, compression, and shear force values computed by a finite element analysis of five limiting-case truss orientations.
Note that the maximum compression assumes an acceleration of 10$g$, which simulates the effect of a shock during site transport, and is intended to be a worst-case scenario.
We used these values as the strut strength requirements.
The full results of this analysis are given in Table~\ref{table:truss_loads}.
}
\label{table:maxTrussLoads}
\end{table}


\begin{figure*}[t]
\centering
\includegraphics[width=0.85\textwidth, trim={2.2cm, 7cm, 4.3cm, 3.4cm},clip]{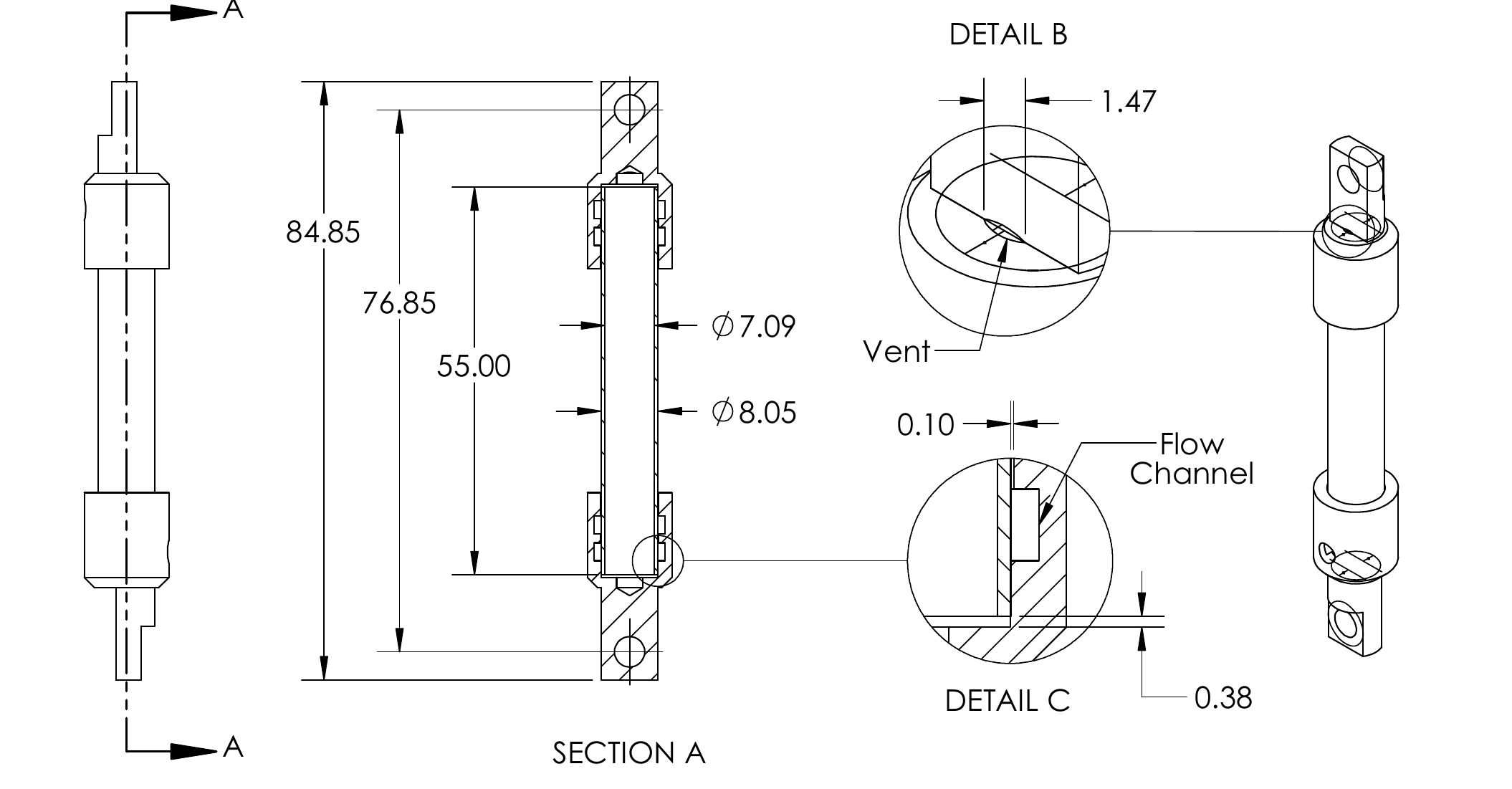}
\caption{
Mechanical drawing of one strut with dimensions.
The units are mm, and the temperature is 300~K.
The truss discussed in this paper (see Figures~\ref{fig:TrussSimOrientation} \& \ref{fig:truss_assembly}) is made with 24 of these struts.
The scale for Detail B and Detail C is 4:1.
Detail C shows one of the two flow channels described in Section~\ref{sec:aluminum_end_caps}.
}
\label{fig:strut_drawing}
\end{figure*}


\begin{figure}
\centering
\includegraphics[width=\columnwidth, trim={6.75cm, 3cm, 6.25cm, 1.9cm}, clip]{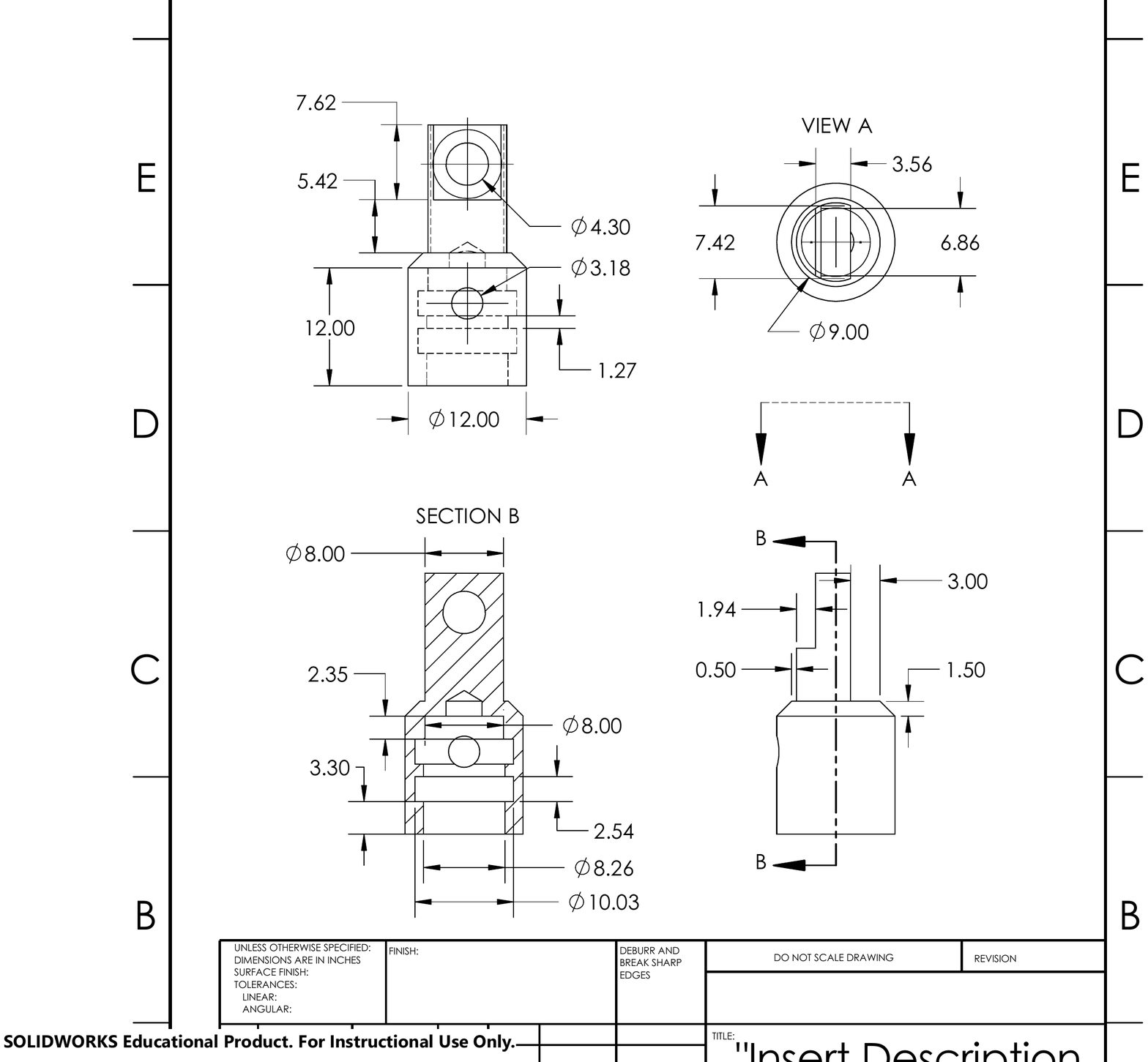}
\caption{
Mechanical drawing of the strut end cap with dimensions.
The units are mm, and the temperature is 300~K.
The 3.18~mm hole shown in the left two panels is the injection point for the epoxy adhesive.
The fit between the carbon fiber tube and the end cap at the back of the socket is designed to be tight to help with alignment and to prevent epoxy adhesive from clogging the vent (see Figure~\ref{fig:strut_drawing}).
The diameter of the back of the socket is shown as 8.00~mm in the drawing, but it is actually tailored to fit the measured diameter of the carbon fiber tube, which varies because of manufacturing tolerances.
}
\label{fig:end_cap_drawing}
\end{figure}


Figure~\ref{fig:TrussSimOrientation} shows the simulation setup for configurations \#3 (panel $a$) and \#5 (panel $b$), which are the truss orientations that exhibit the highest individual strut loads.
In all simulations, the assembly is fixed at the surface of the 4~K ring that mates to the 4~K stage of the cryostat.
This surface is colored green in the figure.
The load on the truss is represented by a point mass whose magnitude and location are equal to the center of mass of all $\leq$ 1~K instrument components.
The point mass is rigidly attached to the 1~K ring of the truss in the simulation. 

To compute the loads on the individual struts, the free-body force $\vec{F}$ on each strut was extracted from the results of the five different simulations.
The free-body force was then decomposed into its axial $F_a$ and radial $F_r$ force components:
\begin{equation}
F_a = \vec{F} \cdot \hat{a}
\end{equation}
and
\begin{equation}
    F_r = \sqrt{ | \vec{F} |^2 - F_a^2}
\end{equation}
where $\hat{a}$ is a unit vector pointing along the length of the strut.
The full results of this analysis are given in Table~\ref{table:truss_loads}, but the maximum compression, tension, and shear values are reported here in Table~\ref{table:maxTrussLoads} for convenience.
These maximum values serve as the performance requirements for the struts.


\subsection{Strut Design}
\label{sec:strut_design}

The strut consists of a central carbon fiber tube (Section~\ref{sec:cf_tube}) fitted with two aluminum 6061 end caps (Section~\ref{sec:end_cap}).
The end caps are affixed to the carbon fiber tube with epoxy adhesive (Section~\ref{sec:epoxy_joint}).
Mechanical drawings of the strut are shown in Figure~\ref{fig:strut_drawing}.


\subsubsection{Aluminum End Cap}
\label{sec:end_cap}

Mechanical drawings of the end cap are shown in Figure~\ref{fig:end_cap_drawing}.
Each aluminum end cap consists of a socket, which is the interface between the end cap and the carbon fiber tube, and a tab, which is the interface between the strut and the truss.
Two cylindrical flow channels were machined into the interior of the end cap sockets with a Woodruff keyseat cutter to increase the strength of the epoxy joints (more in Section~\ref{sec:epoxy_joint}).
A small hole was drilled through one side of the end cap so epoxy can be injected into the flow channels after the carbon fiber tube is inserted. 
Another hole was drilled into the closed end of the end cap so the interior volume of the carbon fiber tube can vent.
Finally, the diameter of the socket at the closed end, beyond the flow channels, is slightly smaller than the diameter at the open end.
This feature results in a tighter fit of the end cap around the carbon fiber tube, which helps with alignment, and it also prevents epoxy adhesive from flowing towards the closed end of the end cap and clogging the vent hole during epoxy injection.


\begin{figure}[t]
\centering
\includegraphics[width=\columnwidth]{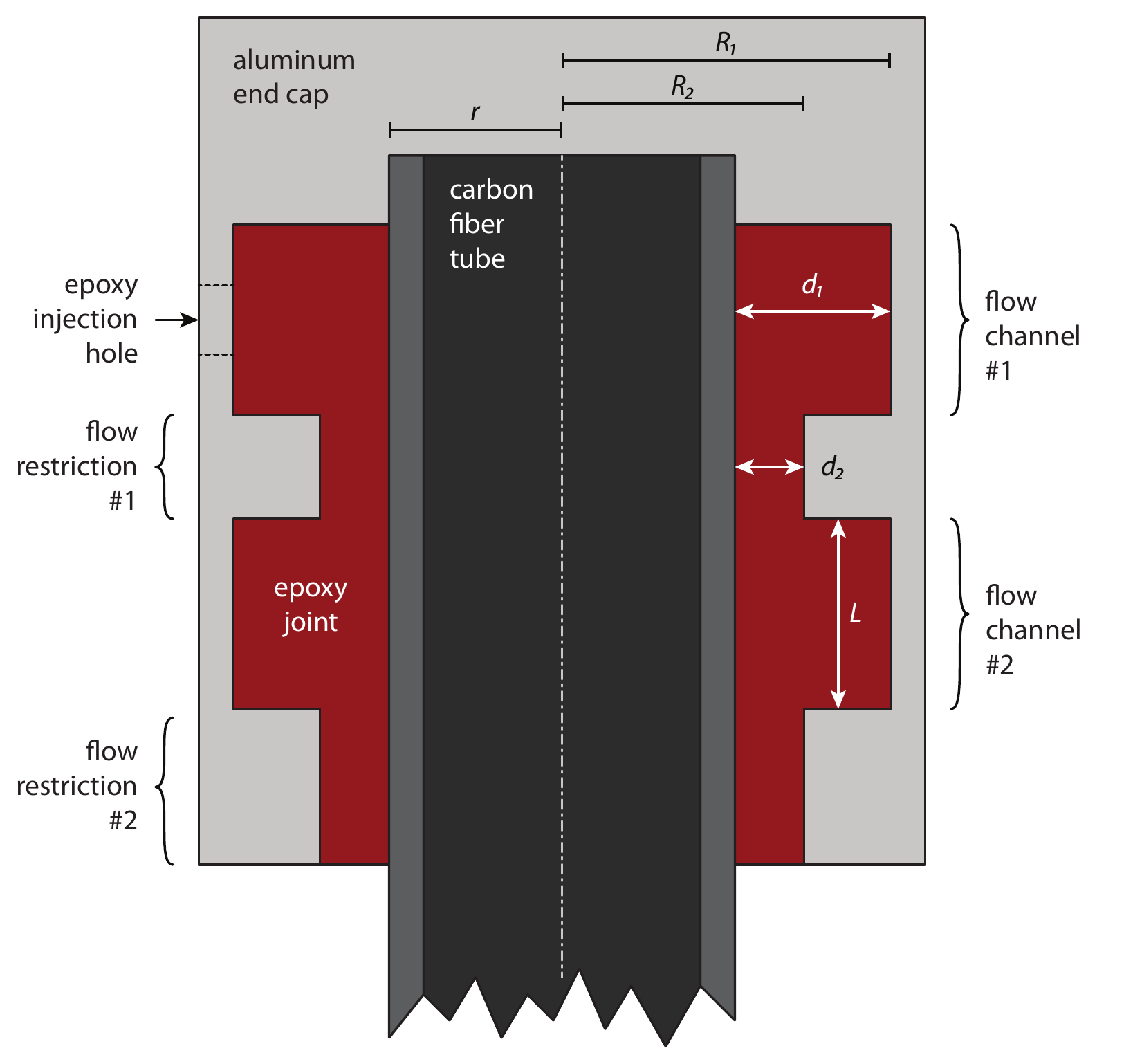}
\caption{
End cap schematic highlighting the epoxy joint.
Each end cap has two identical flow channels.
The actual end cap dimensions are given in Figure~\ref{fig:end_cap_drawing}. 
See Section~\ref{sec:epoxy_joint} for more detail.
}
\label{fig:end_cap_schematic}
\end{figure}


\subsubsection{Epoxy Adhesive Joint}
\label{sec:epoxy_joint}

A schematic of the epoxy joint that forms inside the end cap after the epoxy adhesive hardens is shown in Figure~\ref{fig:end_cap_schematic}.
The bond between the epoxy adhesive and the carbon fiber tube should be very strong because the resin in the carbon fiber tube is also epoxy.
After hardening, the carbon fiber tube and the epoxy joint are effectively one part, and the carbon fiber tube can only be pulled out of the aluminum end cap if the epoxy joint cohesively fails or if the aluminum socket fails.
It is important to note that with this approach, we are not relying on the strength of the adhesion at the aluminum/epoxy interface because the carbon fiber tube is mechanically captured inside the end cap after hardening.
Even if the epoxy fully deadheres from the aluminum, the carbon fiber tube will still be captured in the end cap socket and the strut should not fail.
Our main design task focused on selecting the dimensions of the features in the end cap socket so the epoxy remains in the elastic limit at both room temperature and at approximately 1~K.
If the strain in the epoxy joint is too great, then the epoxy joint could crack, which could lead to strut failure.

We did a series of calculations to determine the dimensions of the features in the end cap socket.
These calculations assume any thermal contraction in the carbon fiber tube is negligible when compared with the thermal contraction of the aluminum and the epoxy adhesive~\cite{reed_1994}.

In our calculations, the radius of the end cap socket is $R$ and the outer radius of the epoxy joint is $J = r + d$, where $r$ is the radius of the carbon fiber tube, and $d$ is the thickness of the epoxy joint. 
At room temperature, $R = J$.
At cryogenic temperatures (either 1~K or 4~K in the truss), the aluminum and the epoxy joint will thermally contract.
After contraction, the various dimensions become
\begin{align}
R^{\prime} =& ~ R \, ( 1 - \alpha_a ) \\
d^{\prime} =& ~ d \, ( 1 - \alpha_e ) \\
r^{\prime} =& ~ r \\
J^{\prime} =& ~ r^{\prime} + d^{\prime}.
\end{align}
In these equations, $\alpha_e$ and $\alpha_a$ are the integrated thermal contraction coefficients for the epoxy adhesive and for aluminum 6061, respectively.
We chose to use 3M Scotch-Weld 2216 for the epoxy adhesive, which has been studied for use in cryogenic applications~\cite{3m_2018, amils_2016, rondeaux_2002}.
The coefficient of thermal expansion is 102~$\mu$m/m/K between 273 and 313~K, and 45~$\mu$m/m/K between 173 and 273~K~\cite{cote_2011}.
Using these measurements, we assumed $\alpha_e = 1.5 \times 10^{-2}$ between 1 and 300~K.
This assumption is conservative because the measurements suggest the contraction is slowing as temperature decreases, but we are assuming linear contraction from 173~K down to 1~K.
For the aluminum, we used $\alpha_a = 4.62 \times 10^{-3}$~\cite{weisend_1998}.

At cryogenic temperatures, $R^{\prime}$ will not necessarily be equal to $J^{\prime}$.
This fact indicates the epoxy joint is stressed into compliance, and it is likely under tension or compression.
The associated strain in the epoxy joint is
\begin{align}
\varepsilon = \frac{R^{\prime}-J^{\prime}}{d} = \alpha_e - \frac{R}{d} \alpha_a.
\end{align}
If $R^{\prime}~>~J^{\prime}$ then $\varepsilon$ is positive and the epoxy joint is under tension.
If $R^{\prime}~<~J^{\prime}$ then $\varepsilon$ is negative and the epoxy joint is under compression.
Interestingly, it is possible for $\varepsilon$ to be zero, in which case the epoxy joint would experience neither tension nor compression, just as it did at room temperature.
Looking at Figure~\ref{fig:end_cap_schematic}, for the flow channels in the end cap, $d = d_1$ and $R = R_1$; elsewhere, $d = d_2$ and $R = R_2$.
Given the chosen socket dimensions (see Figure~\ref{fig:end_cap_drawing}), at 1~K, we expect $\varepsilon_1 = -8.4 \times 10^{-3}$ in the flow channels (slight compression) and $\varepsilon_2 = -0.17$ elsewhere (strong compression).
Here, $\varepsilon_1$ is computed with $R_1$ and $d_1$, while $\varepsilon_2$ is computed with $R_2$ and $d_2$.

A similar argument holds for the length of the flow channels, which is dimension $L$ in Figure~\ref{fig:end_cap_schematic}.
In this direction, the strain is
\begin{align}
\varepsilon = \frac{[ \, L (1 - \alpha_a) \, ] - [ \, L (1 - \alpha_e) \, ]}{L} = \alpha_e - \alpha_a .
\end{align}
Since $\alpha_e > \alpha_a$, the epoxy adhesive in the flow channel will always be in tension in the axial direction.
Given the values of $\alpha_e$ and $\alpha_a$ we are using, we expect $\varepsilon$ in this direction to be $1.0 \times 10^{-2}$. 

\begin{figure}[t]

\centering
\includegraphics[width=\columnwidth]{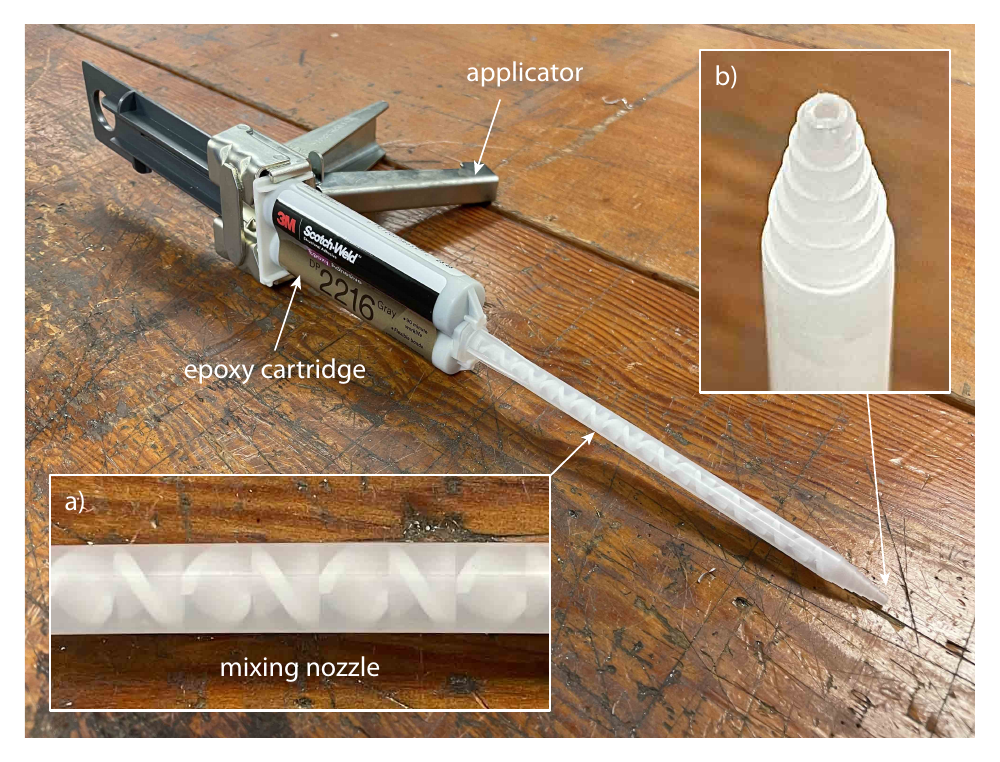}
\caption{
Epoxy adhesive injection system.
The applicator pushes the accelerator and the base (parts A and B) out of the cartridge and into a bayonet style static mixing nozzle with a stepped tip.
Inset $(a)$ shows the elements in the nozzle that mix the two parts.
Inset $(b)$ shows the stepped tip on the nozzle.
The area of the hole in the stepped tip motivated the chosen cross-sectional area of the flow restrictions in the end cap (see Section~\ref{sec:epoxy_joint}).
}
\label{fig:epoxy_adhesive}
\end{figure}

The dimension $d_2$ in Figure~\ref{fig:end_cap_schematic} was chosen to mitigate the formation of voids and allow the epoxy to flow properly during assembly.
Inside the end cap socket there are the two flow channels as well as two flow restriction sections with a radius equal to $r + d_2$.
The epoxy adhesive is injected through a hole in the side of end cap that leads to flow channel~\#1.
The epoxy injection hole is shown in Figures~\ref{fig:strut_drawing}, \ref{fig:end_cap_drawing}, and \ref{fig:end_cap_schematic}.
During injection, the epoxy adhesive will flow where the cross-sectional area is largest, which is into flow channel~\#1.
After flow channel \#1 is full, the epoxy adhesive is pushed through flow restriction \#1 and starts to fill flow channel \#2.
Now, flow restriction \#2 ensures flow channel \#2 fills properly.
The socket cavity is assumed full when epoxy adhesive starts to emerge from the bottom of flow restriction \#2.
Without the flow restrictions, the flow channels may not fill completely leaving air pockets that might compromise the performance of the strut.
The specific $d_2$ dimension was chosen so the cross-sectional area of the flow restriction is slightly larger than the area of the epoxy adhesive injection/mixing nozzle (see Figure~\ref{fig:epoxy_adhesive}), which helps encourage flow.
If $d_2$ is too small, then it is hard to push any epoxy past flow channel \#1.


\subsubsection{Carbon Fiber Tube}
\label{sec:cf_tube}

Three different carbon fiber tube variants were used in the strut samples in this study.
Two of the three variants were sourced by Clearwater Composites\footnote{Clearwater Composites, LLC., 4429 Venture Avenue, Duluth, MN 55811, USA}, and they are referred to as CW1 and CW2 throughout this paper.
The CW1 tubes\footnote{P/N: 0.250-0.320-TW has ID = 0.250~in and OD = 0.320~in} have an 8.13~mm outer diameter (OD) and a 6.35~mm inner diameter (ID).
These tubes are fabricated with multiple layers of high-strength unidirectional carbon fiber prepreg\footnote{Prepreg is the common term for fibers that are pre-impregnated with a resin system that includes the proper curing agent.}, and are wrapped with carbon fiber fabric.
The carbon fiber fabric is woven into a $2 \times 2$ (two over, two under) twill pattern and there are 3,000 filaments in each tow (or bundle).
The CW2 tubes\footnote{P/N: 0.279-0.317-TWA has ID = 0.279~in and OD = 0.317~in} are similar to the CW1 tubes, except they have an 8.05~mm OD, a 7.09~mm ID, and they lack the prepreg contained in the CW1 tubes.
The carbon fiber prepreg in the CW2 tubes was removed to reduce the cross-sectional area of the tube wall.
This change was made to improve the thermal properties of the CW2 tubes at the expense of mechanical strength.
The third carbon fiber tube variant is manufactured by vDijk Pultrusion Products\footnote{vDijk Pultrusion Products, Aphroditestraat 24, NL-5047 TW TILBURG, The Netherlands} and made with a pultrusion process.
We refer to these tubes as DPP throughout the paper.
The DPP tubes\footnote{Item \#: CS-529 2M at The Composite Store, Inc.} have an 8.00~mm OD and a 7.00~mm ID.
They feature straight Torayca T700 carbon fibers\footnote{Toraycma, 19002 50th Avenue East, Tacoma, WA 98446, USA} cured in high-temperature Bisphenol~A epoxy resin, and there is no fabric wrap.
These tubes are easily deformed with radial finger pressure.


\begin{figure*}[t]
\centering
\includegraphics[width=0.7\textwidth]{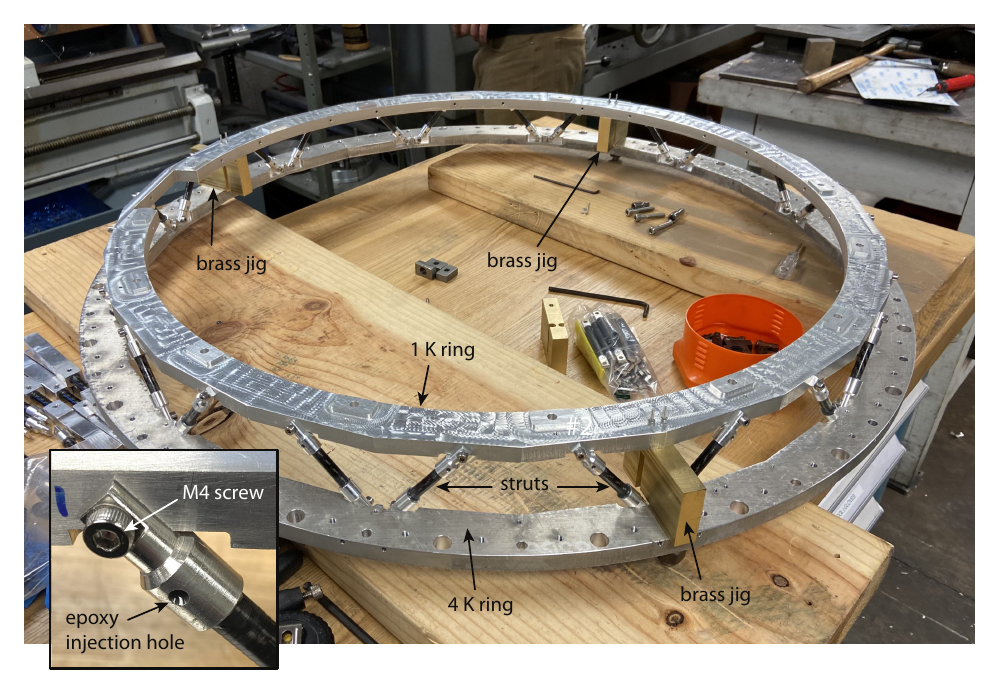}
\caption{
A photograph of one truss during assembly.
The 4~K ring and the 1~K ring were held in place using three brass jigs.
The carbon fiber tubes and the end caps were then installed.
Finally, the epoxy was injected into the end caps.
The struts in this truss were made using CW2 tubes.
This assembly method ensured the rings were parallel and the strut lengths were precisely correct.
The inset shows a close-up view of one end cap.
The assembled truss was strength tested, and the test results are given in Section~\ref{sec:truss_testing}. 
For scale, the inner diameter of the 1~K ring is 68~cm.
}
\label{fig:truss_assembly}
\end{figure*}


\subsection{Strut Assembly Procedure}
\label{sec:strut_assembly_procedure}


%
The length of the bond section at the end of each carbon fiber tube was abraded using a wet-sanding process.
We used 600 grit SiC wet/dry paper for the DPP tubes and a 3M 7447 Scotch-Brite pad for the CW1 and CW2 tubes.
To check that the hydrophobic outer barrier of the carbon fiber was removed after sanding, we used a standard water-break test.
Afterward, most of the water was removed with a compressed air blast and then we allowed the tubes to open air dry for at least 30 minutes before proceeding with the assembly. 

The end cap sockets were cleaned in an ultrasonic bath with a mild detergent for 30 minutes, and then thoroughly rinsed and dried using compressed air.
The bonding surfaces of the sockets were abraded with a 9.5~mm stainless steel tube brush\footnote{McMaster-Carr P/N:~4810A22} held in a battery operated hand drill run on high speed in both directions for several seconds.
This brush was dipped in isopropyl alcohol before and after each socket.
Any remaining alcohol on the bonding surface was removed with compressed air, and the end caps were allowed to sit for an additional 10 minutes before proceeding with the strut assembly. 

The tube-socket assemblies were then secured using fixtures that provided the appropriate alignment and spacing of the end caps (see Figure~\ref{fig:truss_assembly} for the truss and Figure~\ref{fig:struts_after} later in Section~\ref{sec:results} for the strut samples).
Within 90~minutes of the surface preparation the joints were bonded with Scotch-Weld 2216 Grey epoxy adhesive using the mixing nozzle/injection method through the 3.18~mm hole in the socket (see Figure~\ref{fig:end_cap_drawing}).
Adhesive was injected until it flowed out from between the socket and the tube around the entire tube surface. 
The nominal cure time for the epoxy is seven~days at 300~K, but this decreases to 120~minutes at 339~K.
The strut samples were cured on the fixture at room temperature for 65~hours then in an oven at 339~K for 120~minutes.
After cooling, they were removed from the fixture.
The struts on the truss shown in Figure~\ref{fig:truss_assembly} were allowed to cure for seven~days at 300~K because we wanted to preserve the truss alignment.


\section{Predicted Strength of Strut Components}
\label{sec:calculated_strength}


\begin{figure}[t]
\centering
\includegraphics[width=\columnwidth]{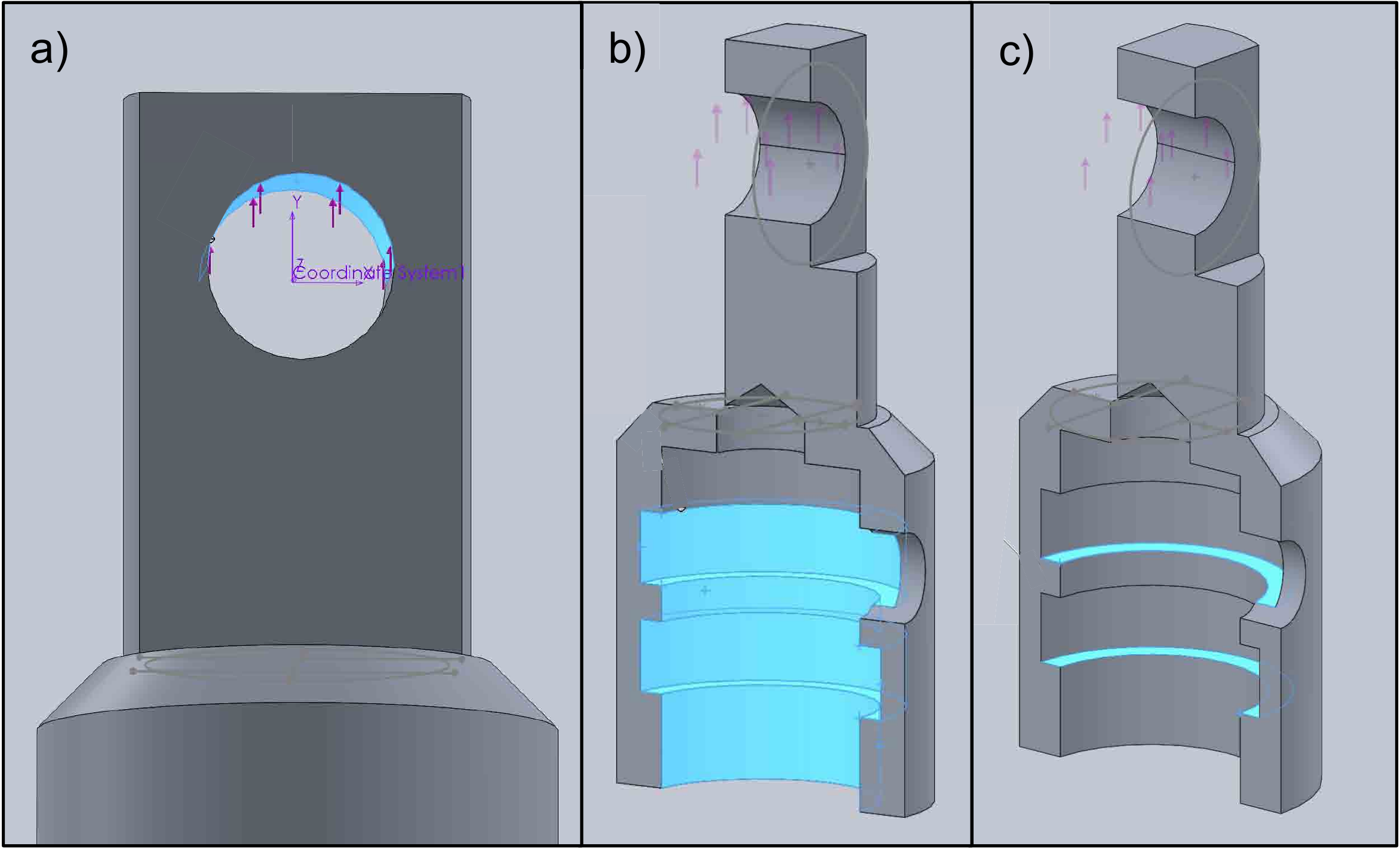}
\caption{
Surfaces used in the FEA of the end cap (highlighted in blue).
A bearing load was applied inside the screw hole in the end cap tab (panel (a)).
To simulate adhesive success, the surfaces shown in panel (b) were fixed.
To simulate adhesive failure, the surfaces shown in panel (c) were fixed.
In the case of adhesive failure, the bond between the epoxy and flow channel walls fails, and the epoxy exerts a normal force on the lower ridge of each flow channel.
}
\label{fig:solidworks-setup}
\end{figure}


\begin{figure*}
\centering
\includegraphics[width=0.95\textwidth]{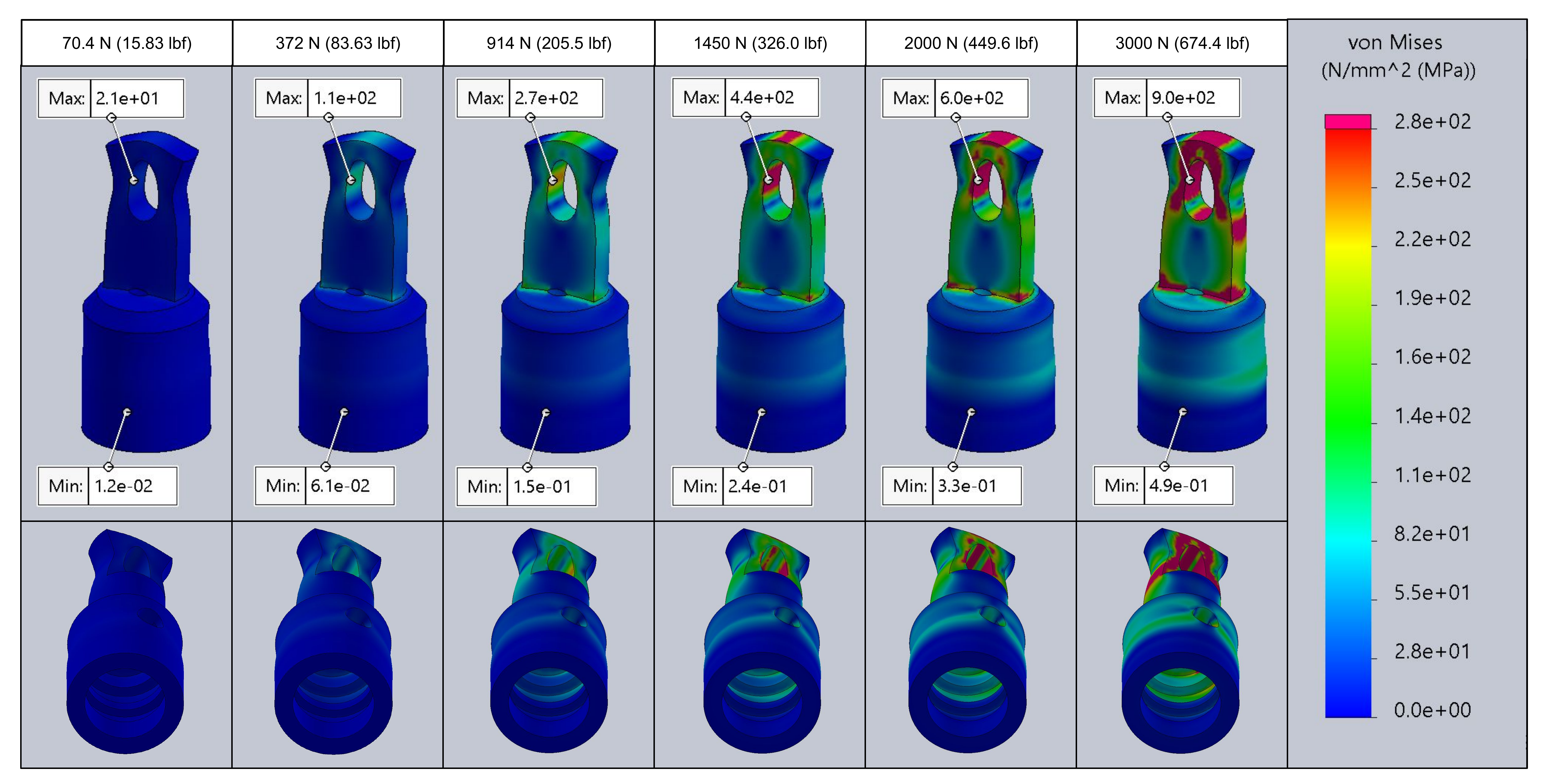}
\caption{
Results of the FEA described in Section~\ref{sec:calculated_strength}.
Here we show the von Mises stress under loads of 70.4, 372, 914, 1450, 2000, and 3000~N.
The minimum and maximum loads expected during observations are 70.4~N and 372~N; the 1450~N load estimates an accidental instrument drop during site transport (see Section~\ref{sec:performance_requirements} and Table~\ref{table:truss_loads}).
At 914~N, the FOS becomes 1 in the end cap near the screw hole (see Figure~\ref{fig:FOS372N}).
The 2~kN and 3~kN loads are useful for understanding the pull-test measurements (see Section~\ref{sec:strut_testing}).
Red indicates where the aluminum is stressed beyond the 275~N/mm$^2$ yield strength.
The scale is capped at the ultimate strength of aluminum, 310 N/mm$^2$; locations where stress exceeds 310 N/mm$^2$ are colored pink.
These plots correspond to simulations conducted in the adhesive failure limit. Adhesive failure allows more deformation in the base and yields a marginally greater von Mises stress.
%
}
\label{fig:vonmises_array}
\end{figure*}


\begin{figure}
\centering
\includegraphics[width=\columnwidth]{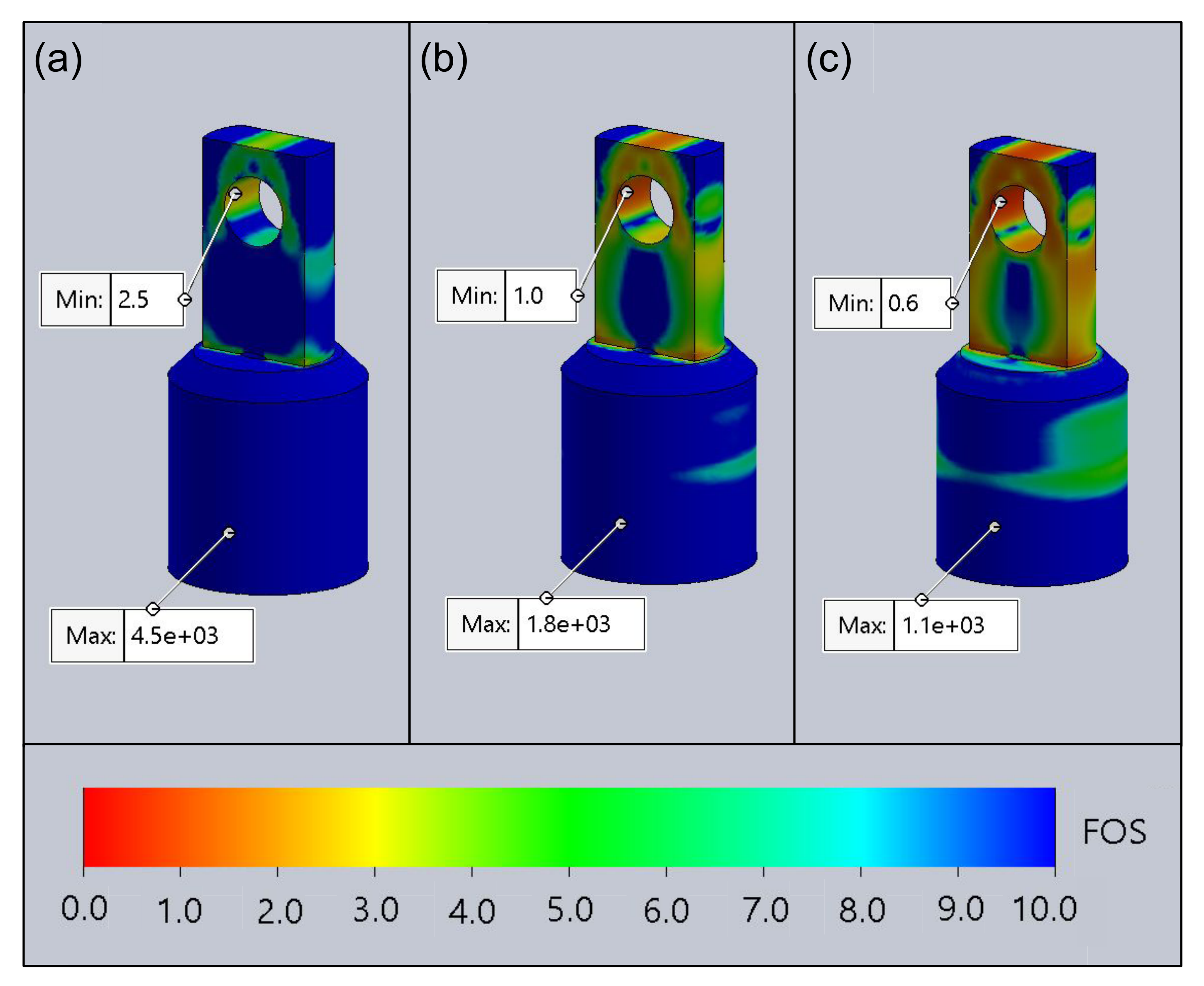}
\caption{
Panel (a): Factor of safety under a load of 372~N.
This is the highest expected load under normal use, according to Table~\ref{table:truss_loads}.
Panel (b): Factor of safety under a load of 914~N.
This is the load where the minimum FOS is 1.
Panel (b): Factor of safety under a load of 1450~N.
This is the load that estimates the effect of an unexpected instrument drop.
FOS is defined as the ratio of the load at which yielding occurs to the highest expected load.
The weakest part of the end cap is inside the screw hole in the tab.
The other weak points are the corners where the end cap tab meets the cylindrical base.
}
\label{fig:FOS372N}
\end{figure}


\begin{table}
\footnotesize
\centering
\renewcommand\arraystretch{1.3}
\begin{tabular}{ccc}
\hline
Property                        & Success       & Failure       \\
\hline
Maximum stress [\,MPa\,]        & 111.582       & 111.585       \\  
Maximum displacement [\,mm\,]   & 0.012         & 0.013         \\ 
Maximum strain                  & 1.27182e-3    & 1.27181e-3    \\ 
Minimum FOS                     & 2.465         & 2.464         \\ 
\hline
\end{tabular}
\caption{
Simulation results for the cases of adhesive success and failure at the maximum expected load of 372~N.
Values for each property are shown to the first digit at which the two cases differ.
Maximum stress refers to the von Mises stress.
}
\label{table:cap_simulation_limits}
\end{table}


\subsection{Aluminum End Caps}
\label{sec:aluminum_end_caps}

We conducted a finite element analysis of the end cap using SolidWorks to identify weak points in the part and to predict the yield strength.
The part is expected to be at least as strong in compression as in tension~\cite{craig_2011}.
Our simulations results were consistent with this expectation, so this section includes the results of the FEA simulations in tension only.
We performed the analysis under two scenarios: when the adhesion between the epoxy and the walls of the end cap socket was successful and when it failed.
In each scenario, we applied a sinusoidally distributed bearing load to the upper surface of the screw hole and fixed the surfaces in the flow channels of the end cap as shown in Figure~\ref{fig:solidworks-setup}.
These two assumptions are intentionally conservative.
In the truss assembly, the compression force provided by the tightened screw will improve the performance by distributing the loading.
The FEA effectively shows what will happen if the screws are loose, which is a worst-case scenario.
We chose to fix the surfaces in the flow channels because we expect more distortion near the screw hole; if we instead fixed the screw hole and loaded the flow channels, deformation would be prevented near the hole, yielding a less realistic result.


Table~\ref{table:cap_simulation_limits} contains the maximum simulated values for stress, strain, and displacement, and the minimum factor of safety (FOS) for each scenario under the maximum expected load of 372~N. 
The differences between the two results are small, and the computed factor of safety is the same to the fourth significant figure, suggesting that in the limit of adhesive failure, the altered force distribution from the epoxy does not make the end cap significantly more likely to fail by yielding.
A major advantage of this design is that in the event of adhesive failure, the epoxy has hardened around the carbon fiber and into the flow channels, so the end cap can not pull free except in the case of bulk failure of the epoxy.

To estimate the load and location at which the end caps begin to undergo plastic deformation, we applied a series of bearing loads to the inside of the screw hole in the tab of the end cap, ranging from 70.4~N to 3.00~kN as shown in Figure~\ref{fig:solidworks-setup}.
For each load, we identified the locations of greatest stress, compared the maximum modeled von Mises stress to the yield strength, and identified the minimum factor of safety in the part.

Figure~\ref{fig:vonmises_array} shows the von Mises stress over the entire end cap when this range of bearing loads is applied. 
The von Mises stress is useful because it characterizes a 3-dimensional stress scenario in a ductile material, as a single quantity representing the energy per unit volume in the part due to  internal distortion under stress.
The commonly used distortion-energy theory of failure predicts that a part will exit the elastic regime and undergo permanent deformation when the von Mises stress exceeds the yield strength of the material (see Reference~\cite{craig_2011} for more detail).
The von Mises stress $\sigma'$ is defined as
\begin{equation}
    \begin{aligned}
    \sigma' =& \frac{1}{\sqrt{2}} [(\sigma_x - \sigma_y)^2 + (\sigma_y - \sigma_z)^2 + (\sigma_z - \sigma_x)^2 \\
     &+ 6(\tau_{xy}^2 + \tau_{yz}^2 + \tau_{zx}^2)]^\frac{1}{2} 
    \end{aligned}
    \label{eq:vonmises-xyz}
\end{equation}
where $\sigma_x$, $\sigma_y$, and $\sigma_z$ represent the tensile stresses applied along the principal axes, and $\tau_{xy}$, $\tau_{yz}$, and $\tau_{zx}$ represent the shear stresses applied in the plane of the first subscript along the direction of the second subscript.
Figure~\ref{fig:vonmises_array} shows that the maximum von Mises stress anywhere in the part remains well below the yield strength of aluminum for the expected loads found in Table~\ref{table:truss_loads}. 
It also shows that yielding first appears around 914~N.

Figure~\ref{fig:FOS372N} shows the FOS over the whole end cap with a 372~N load applied.
Here, the FOS is defined as the ratio of the load at which the part yields to the maximum expected load; a part is considered safe when the FOS at any location meets or exceeds the chosen minimum FOS.
We use the FEA to predict the FOS distribution over the end cap using the convention that the von Mises stress anywhere in the part represents the maximum expected load in that location under the chosen applied load (in our case, 372~N). 
Given the loads predicted in Table~\ref{table:truss_loads} and the shape and material of the end cap, our FEA shows that the minimum FOS is 2.4 (see Figure~\ref{fig:FOS372N}), so we expect the part to exceed the requirement for this application.
Figures~\ref{fig:vonmises_array} and \ref{fig:FOS372N} reveal the region around the screw hole experiences the greatest stress and has the lowest FOS, so this is where we expect the end cap to yield and/or fail first. 

It is important to note that in theory we expect the part to be even stronger at cryogenic temperatures.
The yield strength of aluminum 6061 is 29\% higher at 4.26 K than at room temperature, so the factor of safety will increase by this same factor~\cite{hickey_1962}.


\begin{table}
\footnotesize
\centering

\renewcommand\arraystretch{1.3}

\begin{tabular}{lcccc}
\hline							
Property                        & Unit      & DPP	                & CW1                   & CW2                   \\
\hline
Outer diameter                  & mm        & 8.00                  & 8.13	                & 8.05                  \\
Inner diameter                  & mm        & 7.00                  & 6.35	                & 7.09		            \\
Wall thickness                  & $\mu$m    & 500.                  & 889.                  & 483.		            \\
Cross-sectional area            & mm$^2$    & 11.8                  & 20.2                  & 11.5                  \\
\hline
Tensile modulus                 & GPa       & 140.	                & 83.4  	            & 57.9	                \\
Tensile strength                & GPa       & 2.50	                & 0.880	                & 0.485	                \\
Compression strength            & GPa       & 1.60	                & N/A	                & N/A	                \\
\hline
Ultimate strength of            & kN        & 18.8                  & N/A                   & N/A		            \\
tube in compression             & lbf       & 4,240                 & N/A	                & N/A		            \\
\hline
Ultimate strength of            & kN        & 29.5                  & 17.8                  & 5.56                 \\
tube in tension                 & lbf       & 6,630                 & 4,000                 & 1,250                 \\
\hline
Ultimate elongation             & \%        & 1.8                   & 1.1                   & 0.84	                \\
\hline
\end{tabular}

\caption{
Carbon fiber tube dimensions and calculated/expected strength.
%
%
}
\label{table:cf_tube_info}
\end{table}


\begin{table}
\centering
\footnotesize

\renewcommand\arraystretch{1.3}

\begin{tabular}{lcccc}
\hline
Property	                    & Unit      & DPP                   & CW1                   & CW2                   \\
\hline
Outer diameter                  & mm	    & 8.26	                & 8.26	                & 8.26	                \\ 
Inner diameter                  & mm	    & 8.00	                & 8.13	                & 8.05                  \\
Thickness                       & $\mu$m    & 254.	                & 127.	                & 203.	                \\
Length                          & mm	    & 12.0	                & 12.0                  & 12.0	                \\
Area $\times 10^{4}$            & m$^2$     & 3.11               	& 3.11                  & 3.11              	\\
\hline
Lap shear (19.2~K)              & MPa       & 16.8                  & 16.8                  & 16.8                  \\
Lap shear (297~K)               & MPa       & 22.1                  & 22.1                  & 22.1                  \\
\hline
Strength (19.2~K)               & kN        & 5.23                  & 5.23                  & 5.23              	\\
Strength (297~K)                & kN        & 6.86                  & 6.86                  & 6.86              	\\
\hline
Strength (19.2~K)               & lbf	    & 1,180                 & 1,180                 & 1,180                 \\
Strength (297~K)                & lbf	    & 1,540                 & 1,540                 & 1,540                 \\
\hline
\end{tabular}

\caption{
Epoxy joint details and calculated/expected strength.
The area used in the strength calculation is the area of the interface between the epoxy and the end cap socket, excluding the flow channels.
Therefore, the calculated strength here is the maximum strength one might expect from a socket without flow channels. 
%
%
}
\label{table:epoxy_info}
\end{table}


\subsection{Carbon Fiber Tubes \& Epoxy Adhesive}
\label{sec:carbon_fiber_tubes_epoxy}

The expected strength of the carbon fiber tubes and the epoxy adhesive was computed using information from the various manufacturers.
vDijk Pultrusion Products provided the tensile modulus, the tensile strength, and the compression strength of the DPP tubes.
Clearwater Composites provided the tensile modulus and the tensile strength of the CW1 and the CW2 tubes.
All of these material property values assume the ambient operating temperature is approximately 300~K.
Using this information and the nominal dimensions of the tubes, we calculated the ultimate strength of the DPP tubes in tension and compression and the ultimate strength of the CW1 and CW2 tubes in tension.
The results of these calculations are given in Table~\ref{table:cf_tube_info}.
3M provides lap shear values for the Scotch-Weld 2216 Gray epoxy adhesive at both 19.2~K and 297~K.
We calculated the expected strength of the epoxy joint at these two temperatures using the lap shear information and the approximate area of the epoxy joint.
The epoxy strength calculation results are shown in Table~\ref{table:epoxy_info}.


\subsection{Stainless Steel Screws}
\label{sec:screws}

The struts are secured to the edge of the truss rings with M4 316 stainless steel screws (see Figure~\ref{fig:truss_assembly}).
At this interface, the screws primarily experience shear stress $\tau$ according to
\begin{equation} \label{eq:shear-stress}
    \tau = \frac{F_{\mathrm{applied}}}{A_{\mathrm{screw}}} = \frac{4 \, F_{\mathrm{applied}}}{\pi \, d_{\mathrm{minor}}^{\,2}} ,
\end{equation}
where $F_{\mathrm{applied}}$ is the shear force applied to the screw by the strut end cap, and $A_{\mathrm{screw}}$ is the cross-sectional area of the screw, calculated using the minor diameter $d_{\mathrm{minor}}$, which is 3.14~mm for an M4 screw.
The minor diameter is the narrowest and therefore weakest part of the screw, so using $d_{\mathrm{minor}}$ in the calculation gives a lower limit~\cite{steeve_2012}.
The shear stresses of interest are the shear yield strength\footnote{Beyond the yield strength, the screw exits the elastic regime and enters the plastic regime, in which deformation is permanent.} $\tau_y$ and the shear ultimate strength\footnote{The ultimate strength corresponds to the peak of the stress-strain curve.} $\tau_u$.
To calculate the corresponding loads that can be applied to the screws, we use the standard approximation predicted by distortion-energy theory~\cite{craig_2011}, $\tau_Y = \sigma_Y /\sqrt{3}$ and $\tau_U = \sigma_U/\sqrt{3}$, where $\sigma_Y$ and $\sigma_U$ are, respectively, the yield and ultimate strength of 316 stainless steel in pure tension at room temperature.
Rearranging Equation~\ref{eq:shear-stress} gives the applied loads required to produce each of the shear stresses of interest:
\begin{equation} \label{eq:yield-force}
    F_y= \tau_y \times \frac{\pi}{4}  \left( d_{\mathrm{minor}} \right)^2,
\end{equation}
and
\begin{equation} \label{eq:ultimate-force}
    F_u= \tau_u \times \frac{\pi}{4}  \left( d_{\mathrm{minor}} \right)^2.
\end{equation}
For 316 stainless steel, we assume $\sigma_Y$~=~410~N/mm$^2$~\cite{moore_2009} and $\sigma_U$~=~550 N/mm$^2$, which is the ultimate strength value provided by SolidWorks.
Therefore, we expect that $F_y = 1.8~\mathrm{kN}$ and $F_u = 2.4~\mathrm{kN}$. 
Again, these are lower limits given our conservative assumptions.
We also expect somewhat better performance at cryogenic temperatures as yield strength tends to increase with decreasing temperature \cite{mchenry_1983}.


\section{Results}
\label{sec:results}

Our testing program included (i) individual strut testing, (ii) assembled truss testing, and (iii) measurements of the thermal conductivity of the carbon fiber tubes.
The results from these tests are reported in Sections~\ref{sec:strut_testing}, \ref{sec:truss_testing}, and \ref{sec:thermal_testing}, respectively.
The individual strut testing program had two main objectives.
First, we wanted to verify that the struts meet the strength requirements (Section~\ref{sec:performance_requirements}) and that the measured performance is repeatable. 
Second, a reasonable match between measurement and theory (see Section~\ref{sec:calculated_strength}) would provide confidence in our end cap design philosophy (Section~\ref{sec:strut_design}).
The assembled truss testing was done to ensure that the deployed SO trusses will meet the strength requirements and behave as expected when the instruments are assembled and operating.
The thermal testing was done to show that the thermal conductivity of the carbon fiber tubes is suitable for this application.
Some ancillary tests are described in~\ref{sec:ancillary_tests}.


\begin{figure}[t]
\centering
\includegraphics[width=\columnwidth]{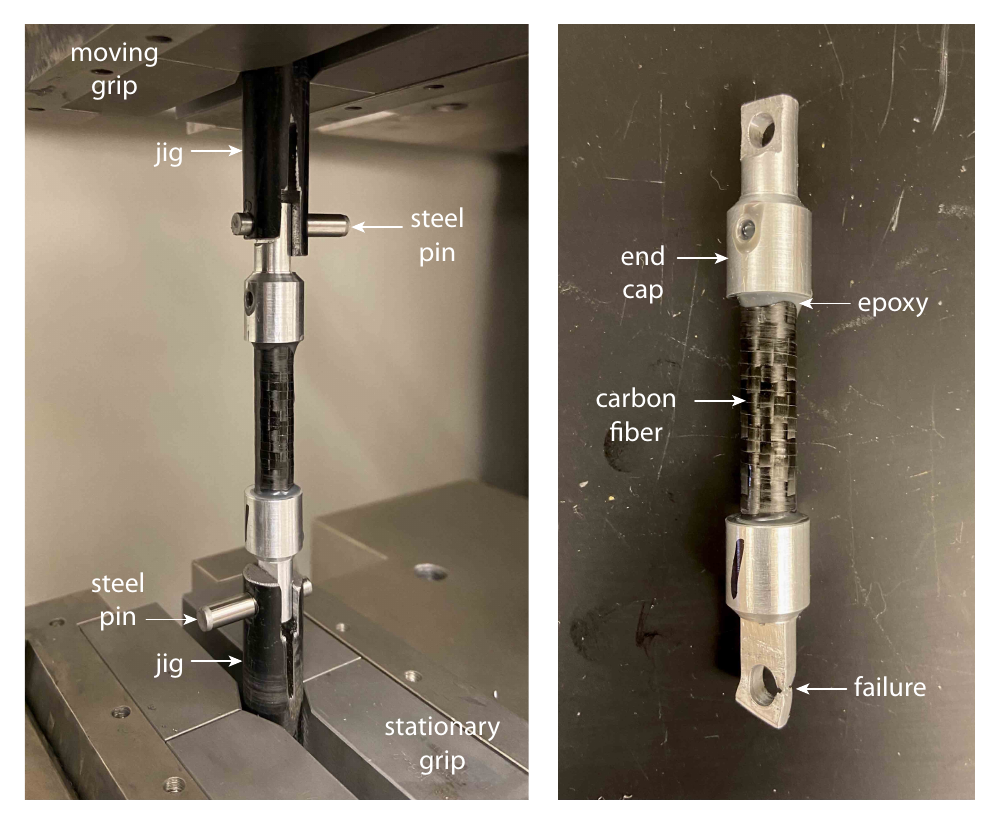}
\caption{
Left: A photograph of one CW2 strut (strut \#~21) mounted in the Instron Model 5985 before testing.
Note that steel pins were used to mount the strut in the test jig because calculations showed that M4 stainless steel screws we use in the truss would fail before any of the other strut elements.
The measurement was therefore focused on testing the aluminum end cap, the epoxy, and the carbon fiber tube. 
The force versus displacement data from this test is plotted in Figure~\ref{fig:pull_test_data}.
Right: A photograph of the same strut after the pull testing finished.
The failure point can be seen in the bottom end cap.
The dimensions of the strut are given in Figures~\ref{fig:strut_drawing}~\&~\ref{fig:end_cap_drawing}.
}
\label{fig:strut_test}
\end{figure}


\begin{figure}[t]
\centering
\includegraphics[width=\columnwidth]{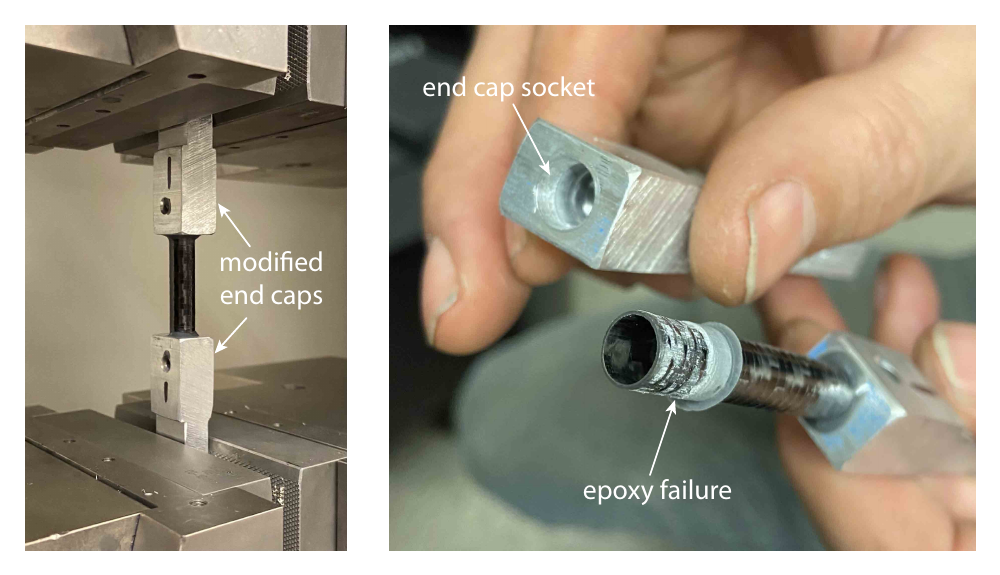}
\caption{
Photographs of the epoxy pull tests.
Left: One of four samples mounted in the Instron Model 5985 before testing. 
This sample has the same carbon fiber tube, epoxy adhesive, and end cap socket as the struts.
The interface between the end cap and the Instron grips, however, was modified to avoid the mounting hole failure (see right panel of Figure~\ref{fig:strut_test}).
This measurement was therefore focused on testing the epoxy strength and the tensile strength of the carbon fiber tube.
The data from this measurement is shown in the lower right panel of Figure~\ref{fig:pull_test_data}.
Right: A photograph of the same strut after the pull testing finished.
}
\label{fig:epoxy_test}
\end{figure}


\subsection{Strut Testing}
\label{sec:strut_testing}


We used an Instron\footnote{Instron, 825 University Ave, Norwood, MA, 02062-2643, USA} Model 5985 to test the strength of the various strut architectures.
The Instron is an electromechanical testing system that applies a tensile or compressive force to a specimen held in a set of grips (see Figure~\ref{fig:strut_test}, for example).
During a test, a force transducer (load cell) measures the applied force, and an extensometer measures changes in the length of the specimen.
This particular Instron model can apply forces up to 250~kN (56,200 lbf).
When a test runs, the Instron records a measurement of the applied force and the associated displacement.
If, for example, the specimen is a solid aluminum cylinder and a tensile test is run, then the resulting measurements can be straightforwardly converted to stress and strain using the cross-sectional area of the specimen.
The slope of the stress data plotted versus the strain data in the elastic limit gives the tensile modulus of the specimen.
Since each strut consists of many components made from different materials with different cross-sectional areas, the conversion to stress and strain is complicated and unnecessary, so we report our results as the applied force versus the measured displacement.


We made a total of 25 struts and four epoxy samples for design verification testing.
The strength of these elements was measured with the Instron during two rounds of testing\footnote{The first round took place on December 14, 2020, and the second round took place on on January 20, 2021.}.
For round one, we prepared the first twenty struts on a custom jig. 
The round-one struts were numbered 1 to 20.
Struts 1 to 10 were made with CW1 tubes and struts 11 to 20 were made with DPP tubes.
For round-two testing we prepared five CW2 struts on the same jig and numbered them 21 to 25.
We also prepared four samples that were specifically designed to allow us to study the strength of the epoxy joint that holds the aluminum end caps to the carbon fiber tubes.
These epoxy test samples had modified end caps, which can be seen in Figure~\ref{fig:epoxy_test}.
In particular, we wanted to see if the flow channels worked as expected (see Section~\ref{sec:strut_design} and Figures~\ref{fig:strut_drawing}~\&~\ref{fig:end_cap_drawing}).
The CW2 tubes are custom-made because the prepreg is removed.
Consequently, the tests were done in two rounds.
The DPP and CW1 tubes were available off-the-shelf, so we started our testing program with these tubes.
The results of the round-one testing advised the round-two testing.


One of our primary concerns is fracture and fatigue in the epoxy joint and/or the carbon fiber tube caused by cryogenic cycling.
To investigate these issues, struts 1 to 5 and struts 11 to 15 were thermally cycled in a liquid nitrogen (LN2) bath five times before testing them in the Instron.
For the thermal cycle we first quickly submerged (shocked) the struts in the LN2 bath and then allowed them to thermalize for approximately three minutes.
We were confident the struts had thermalized because the LN2 first violently boiled after the struts were submerged.
By the end of the three minutes the boiling appreciably quieted down.
After removing the struts from the LN2 bath, we placed them in a thermally-insulating stainless steel bowl full of dry nitrogen boiloff for inspection.
The inspection took approximately four minutes.
We then warmed the struts to room temperature with a heat gun, which took approximately three minutes.
None of the struts broke or showed signs of fatigue during the thermal cycling process.


\begin{figure}[t]
\centering
\includegraphics[width=\columnwidth]{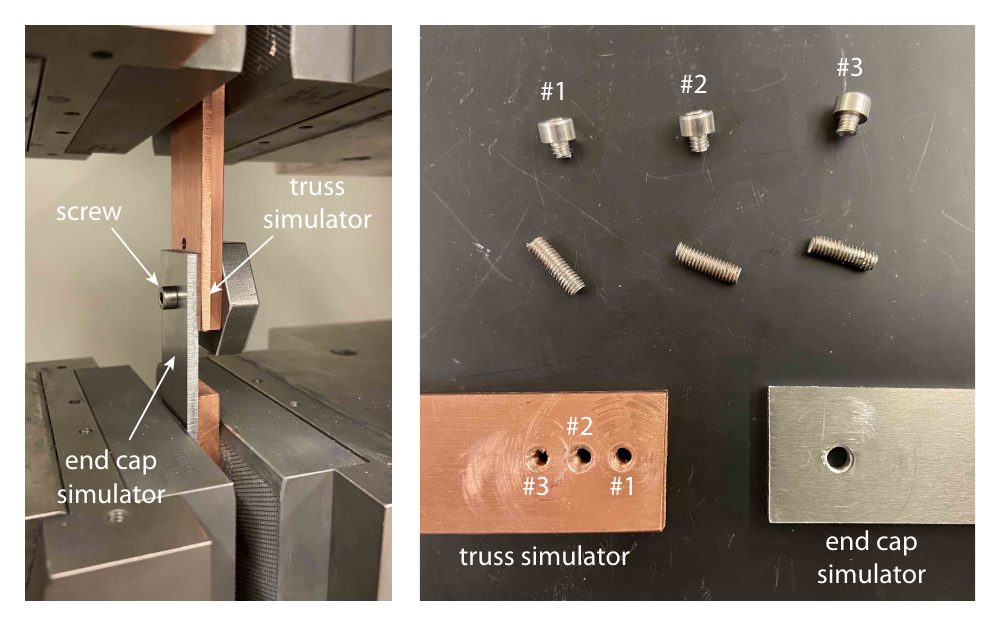}
\caption{
Left: Photo of the pull test setup for the screw testing. 
This setup is designed to simulate the edge of one of the Simons Observatory trusses.
Right: Photo of the three screws after testing.
The screws failed as expected given the shear stress.
A new tapped hole was used for each screw test because we were concerned the threads in the tapped hole would be damaged after the pull test.
The screw numbers in the photo correspond to the curves in Figure~\ref{fig:screw_test_data}.
}
\label{fig:screw_test}
\end{figure}


\begin{figure}[t]
\centering
\includegraphics[width=0.4\textwidth]{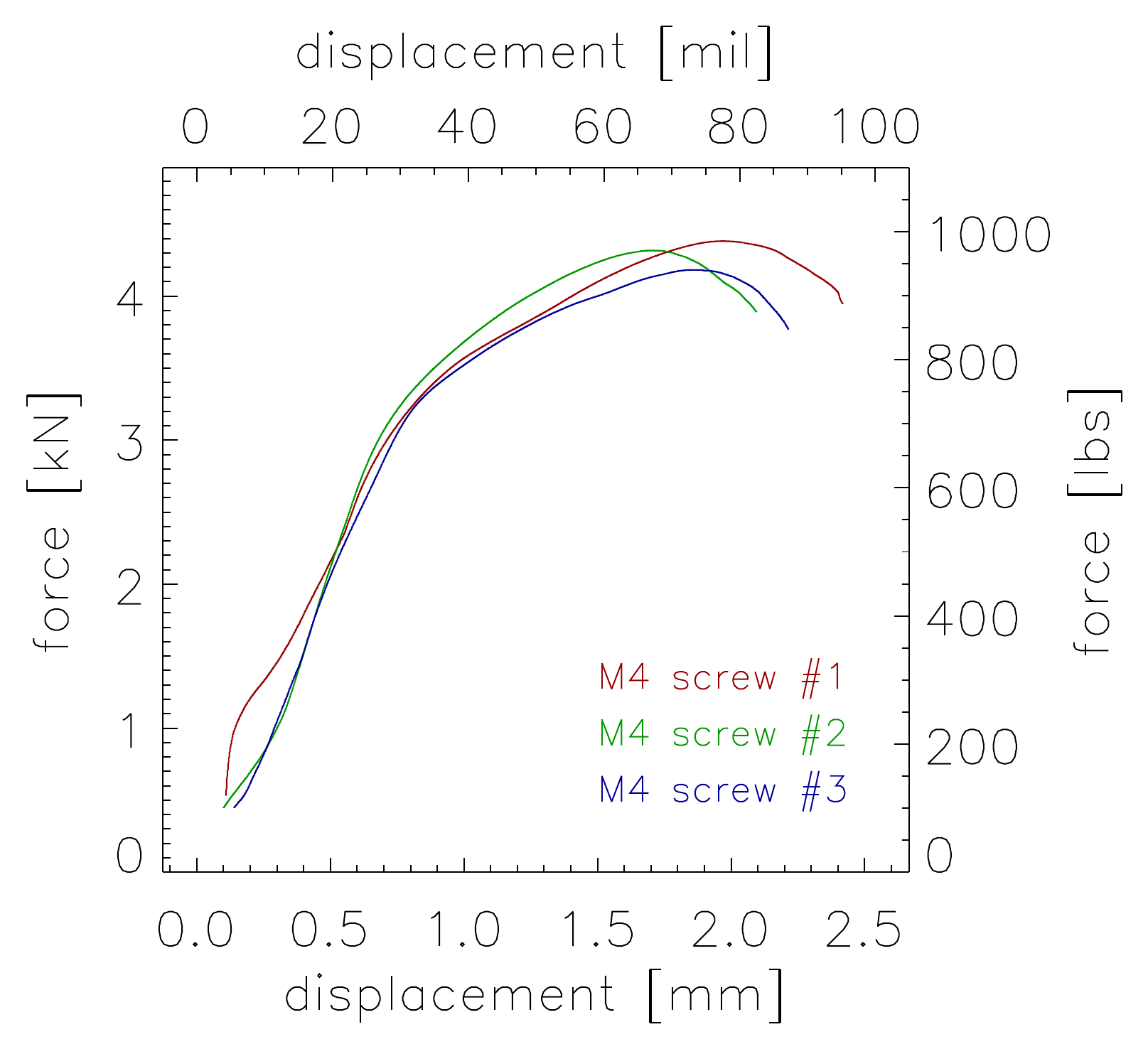}
\caption{
Force versus displacement measurements for the three M4 stainless steel screws that were studied in screw shear test (see Figure~\ref{fig:screw_test}).
These measurements show the screws yield when loaded with approximately 3.0~kN (670~lbf).
}
\label{fig:screw_test_data}
\end{figure}



We wanted the testing program to reveal the strength of the epoxy joint, or at least show that the epoxy joint is not the weakest element in the strut. 
Our initial estimates showed that the screws were expected to be the weakest element in the strut, so we tested them independently in an effort to better test the strength of the epoxy joint.
Given the design of the struts and the truss, the screws experience a shear stress, so we made a jig for the Instron that stresses screws in shear.
A photograph of this test jig is shown in Figure~\ref{fig:screw_test}.
We tested three screws to failure.
The screws failed in shear as expected, and the measurements (see Figure~\ref{fig:screw_test_data}) show the stainless steel yields around 3.0~kN.

\begin{figure*}[p]
\centering
\includegraphics[width=0.4\textwidth]{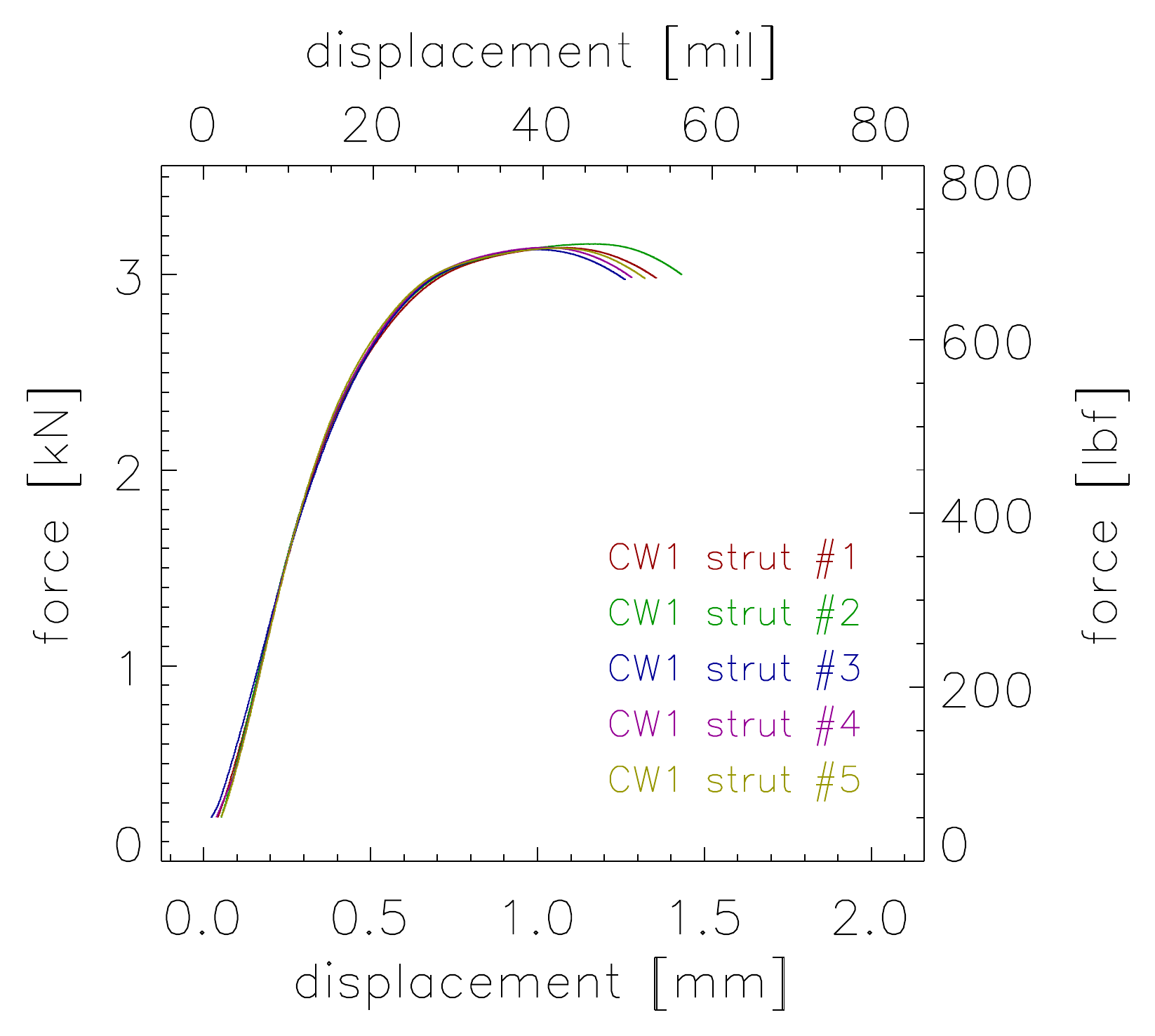}
\hspace{0.1in}
\includegraphics[width=0.4\textwidth]{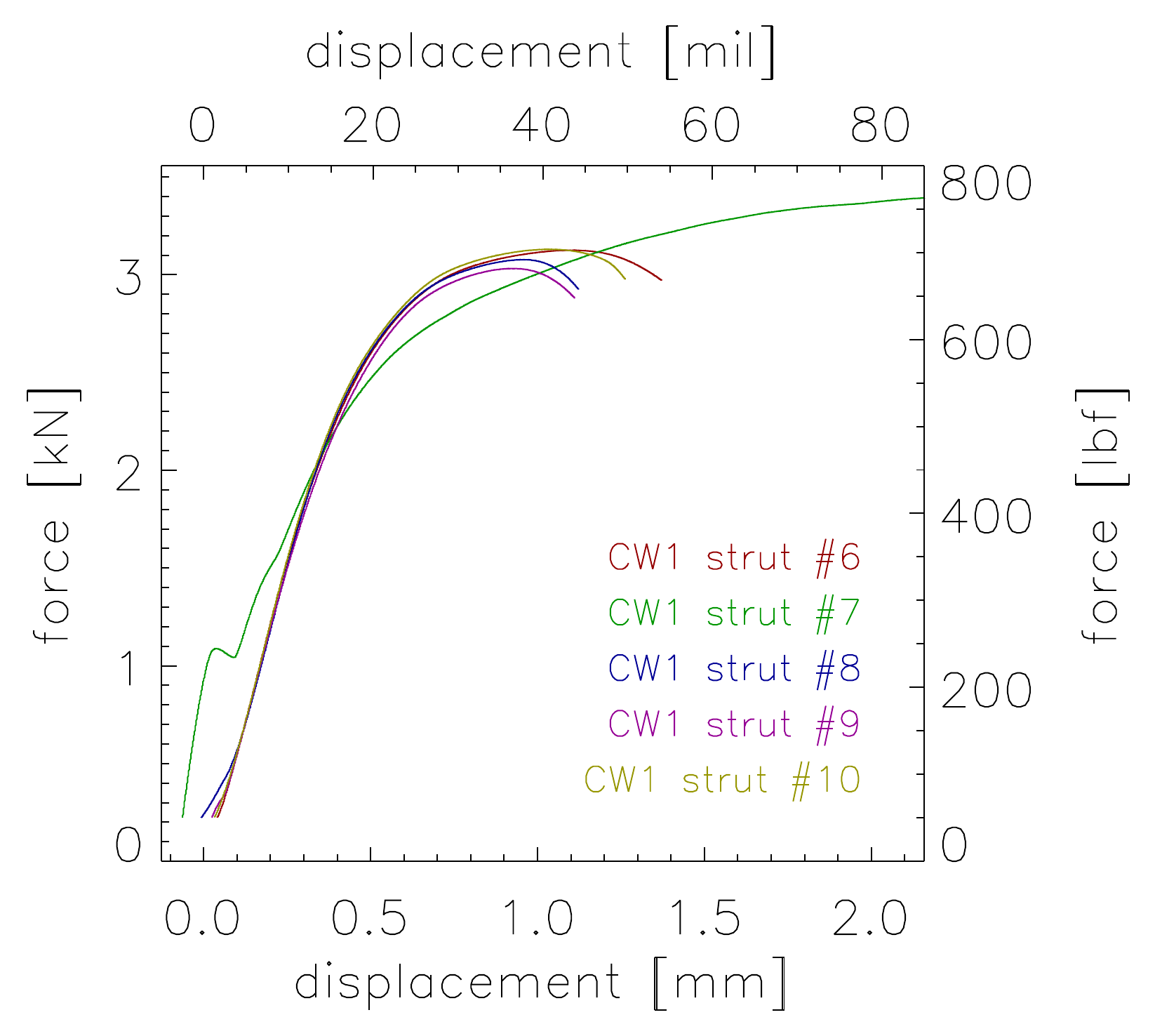}
\includegraphics[width=0.4\textwidth]{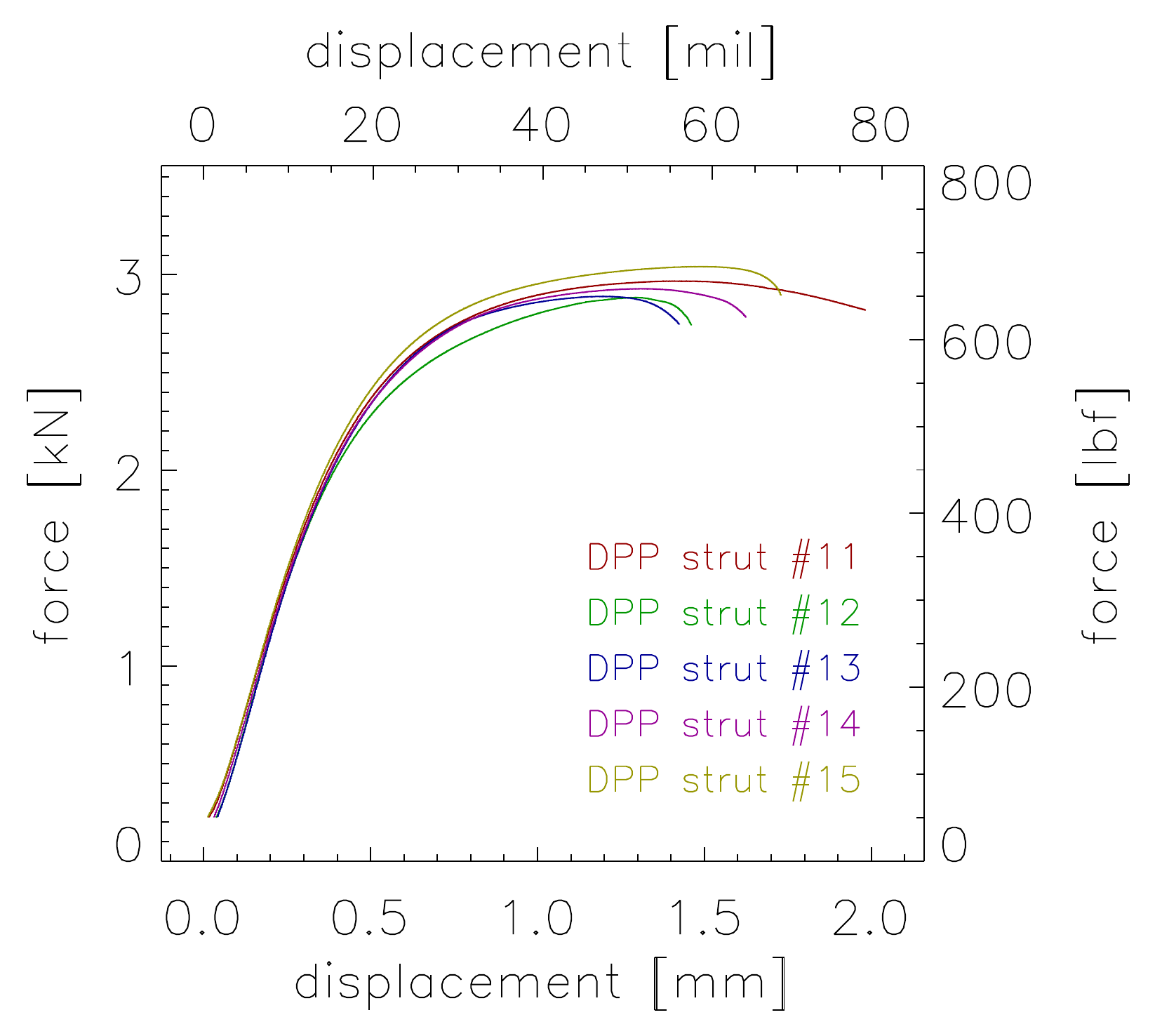}
\hspace{0.1in}
\includegraphics[width=0.4\textwidth]{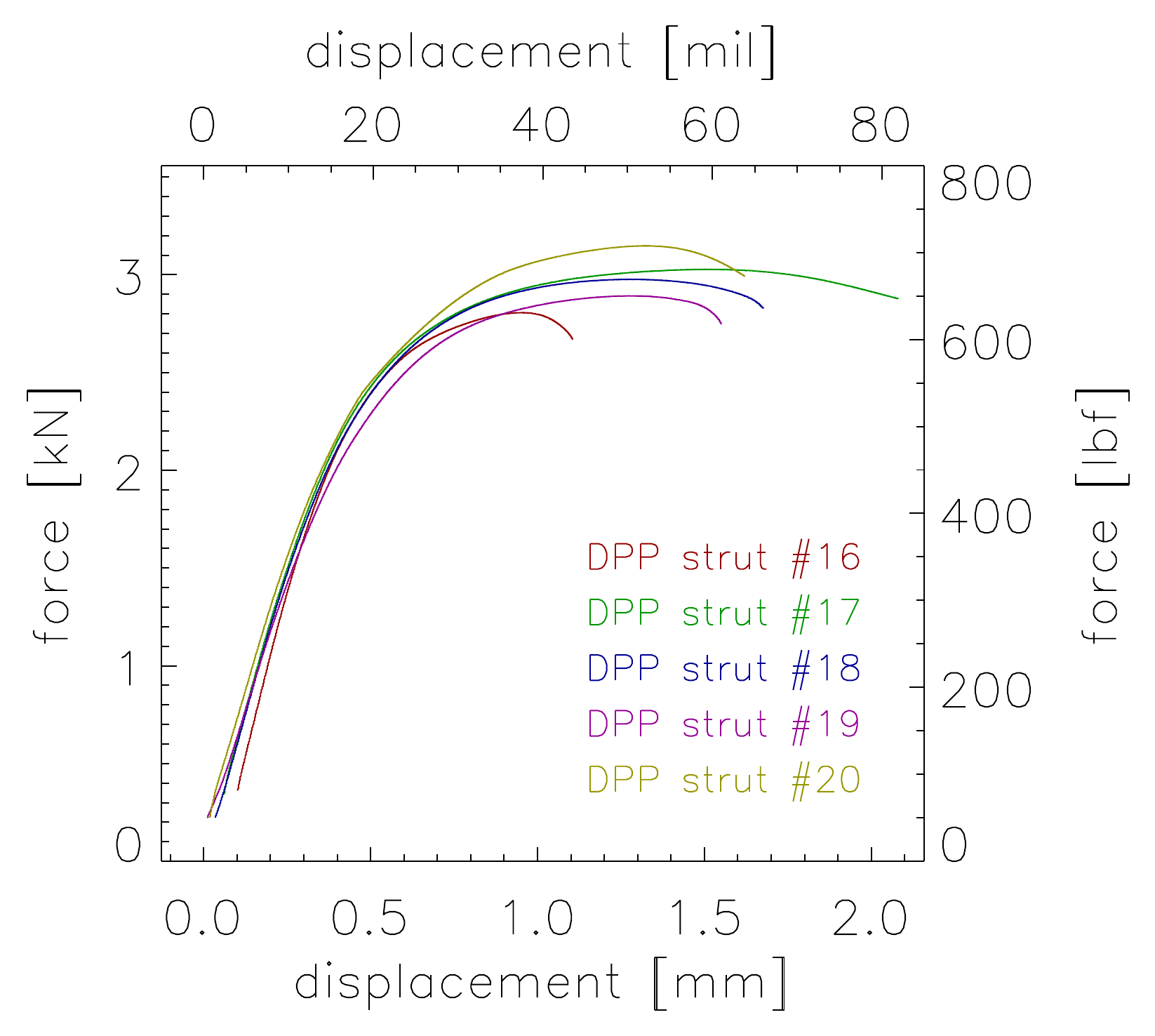}
\includegraphics[width=0.4\textwidth]{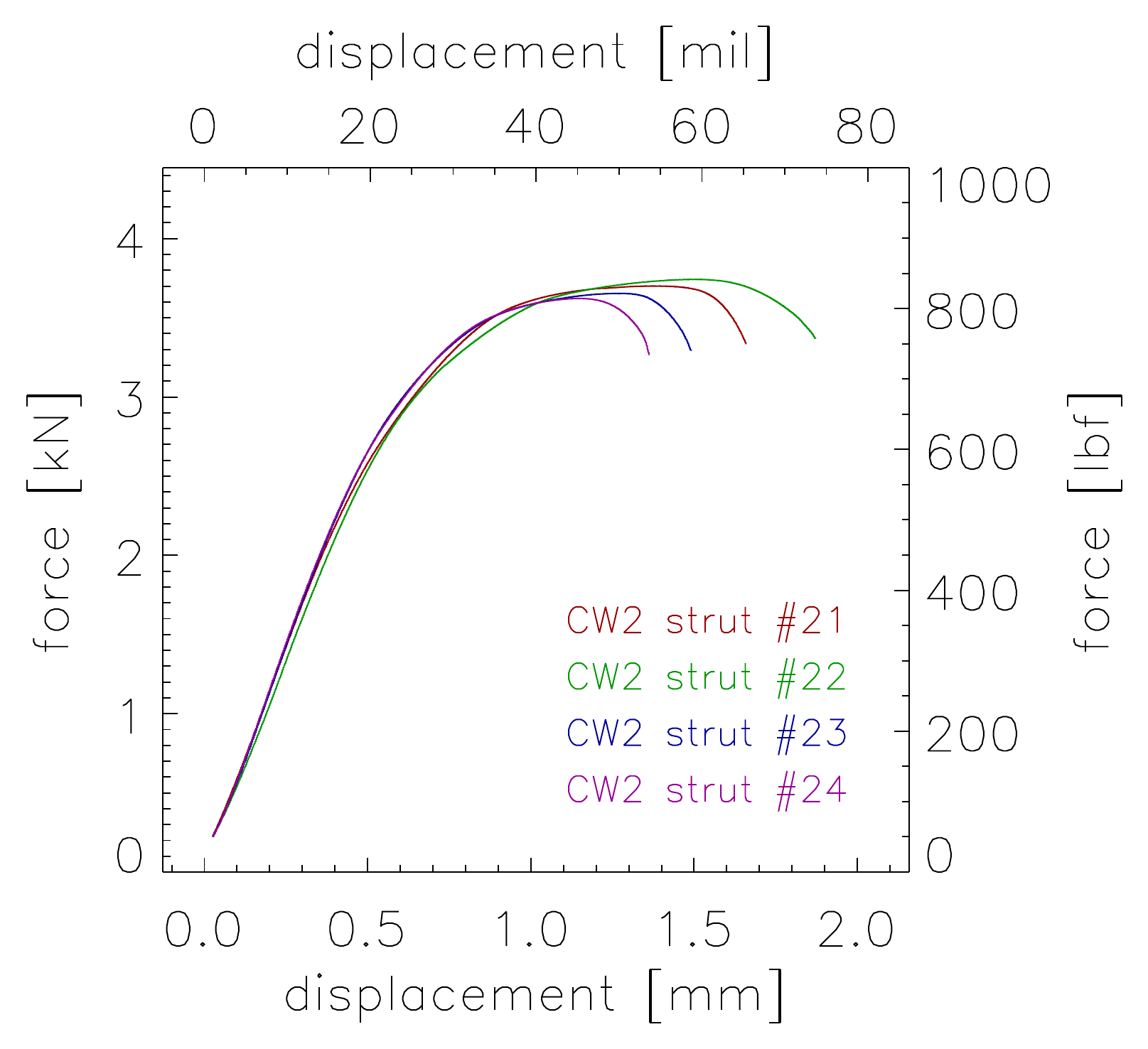}
\hspace{0.1in}
\includegraphics[width=0.4\textwidth]{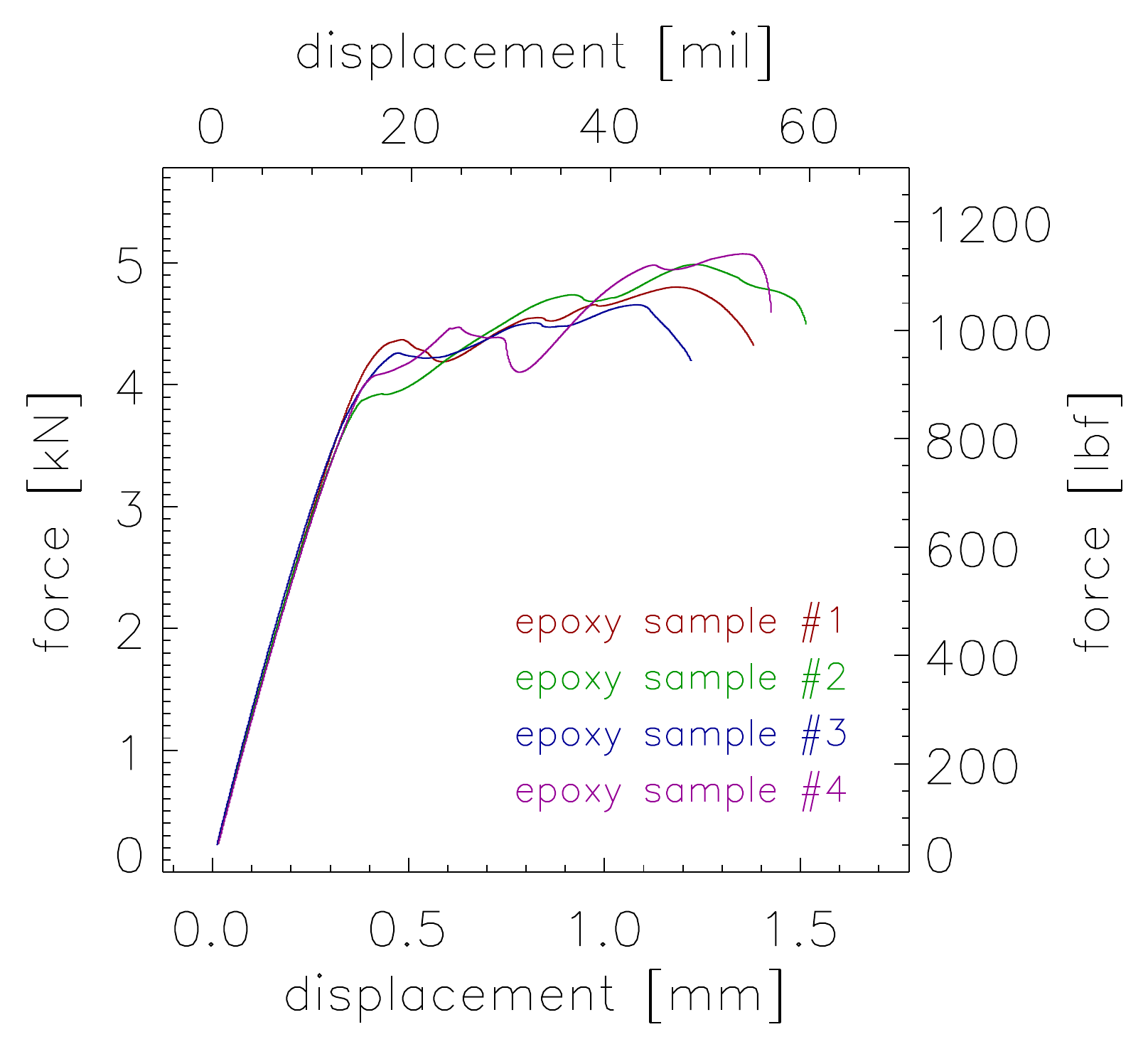}
\caption{
Force versus displacement measurements for the CW1, CW2, and DPP struts as well as the epoxy samples.
Top Row: CW1 results (struts 1-10).
Middle Row: DPP results (struts 11-20).
Top left and center left: cryo-cycled results (struts 1-5 and struts 11-15).
Bottom Left: CW2 results (struts 21-24). See Figure~\ref{fig:strut_test}.
Bottom Right: Epoxy results.
The test setup is shown in Figure~\ref{fig:epoxy_test}, and the expected strength is given in Table~\ref{table:epoxy_info}.
}
\label{fig:pull_test_data}
\end{figure*}
\begin{figure*}[t]
\centering
\includegraphics[width=\textwidth]{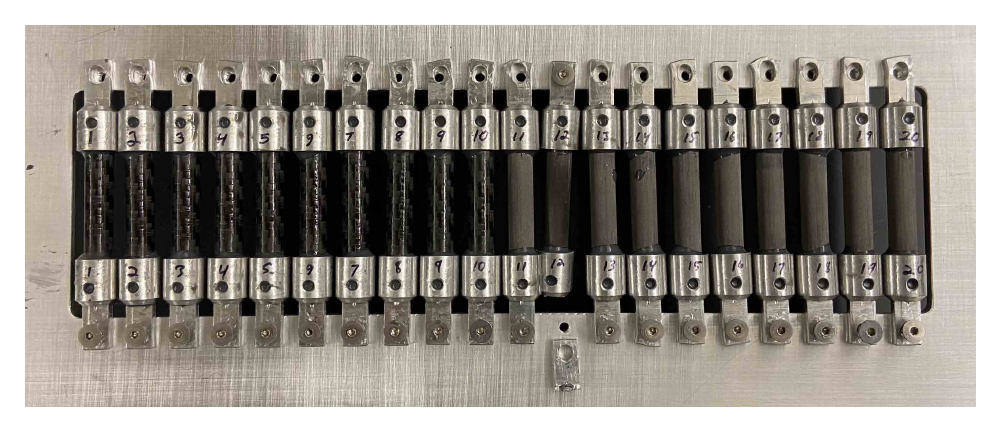}
\caption{
A photograph of the DPP and CW1 struts after the tension testing.
The number written on each strut corresponds to the curve number in Figure~\ref{fig:pull_test_data}.
All struts except strut \#12 failed near the mounting hole for the M4 screw in one of the two the aluminum end cap tabs.
The failure modes observed are similar to the failure modes predicted by the FEA analysis (see Figure~\ref{fig:vonmises_array}).
Struts 1~to~5 and 11~to~15 were thermally cycled to 77~K five times before running the tension test.
%
%
}
\label{fig:struts_after}
\end{figure*}

We tested ten DPP struts, ten CW1 struts, and four CW2 struts to failure in tension using the jig shown in Figure~\ref{fig:strut_test}.
Note that the M4 stainless steel screws were not used in these tests; a stronger steel pin was used instead. 
We also tested the four epoxy samples.
Data from all 28 tests are shown in Figure~\ref{fig:pull_test_data}.
The results were consistent from strut to strut.
The struts that were cryogenically cycled showed the same strength as the struts that were not cryogenically cycled.
A photograph of the CW1 and DPP struts after tension testing is shown in Figure~\ref{fig:struts_after}.


\subsection{Truss Testing}
\label{sec:truss_testing}

After individual struts were shown to have met the necessary strength requirements, the full truss was assembled using the CW2 type carbon fiber. 
Jigs were used to correctly space the two truss rings and the struts were epoxied in place on the jigged truss (see Section~\ref{sec:strut_assembly_procedure} and Figure~\ref{fig:truss_assembly}).
Prior to installation in the SAT, the truss underwent four non-destructive tests to confirm it would meet the loading requirements driven by the different use cases of the fully assembled instrument as described in Section~\ref{sec:performance_requirements}.

The first test simulated stresses caused by differential thermal contraction.
When cooling the SAT from room to operating temperature, the 4~K ring cools faster than the 1~K ring.
This causes differential thermal contraction up to a maximum radial displacement of 1.5~mm, which flexes the struts inward. 
\ref{sec:ancillary_tests} shows that individual struts flexed by this amount remain in the elastic region of the force versus displacement curve.
To test this stress condition on the full truss the 4~K ring was submerged in LN2 until it thermally equilibrated.
The 1~K ring and half the struts remained above the liquid level at a warmer temperature.
The truss was then brought back to room temperature in a bag of nitrogen gas to avoid condensation.
This procedure was repeated five times to mimic repeated cryogenic cycling.
Afterwards the truss was inspected for damage, with particular attention paid to the epoxy joints.
No damage was observed.

The other three tests subjected the truss to compression, tension, and shear stresses and were performed after the LN2 cycling.
To simulate the shear stress (configuration \#3), the truss was placed in an Instron using two flat plates bolted to the truss rings that in turn interfaced to the Instron attachment points.
The 4~K ring remained fixed while a 4.2~kN force was gradually applied to the 1~K ring parallel to the plane of the truss.
This test load is approximately two times the load we expect from the instrument components in the SAT under 1$g$. 
The truss remained in the elastic region of the measured force versus displacement curve and showed no visual signs of damage.
The Instron could not be used for the compression and pull tests as the truss was too wide to fit in the machine.
Instead, weights were used to reach the desired test load levels.
For the compression test (configuration \#5), weights totaling 10.5~kN were stacked on top of the metal plate attached to the 1~K ring of the truss.
This test load is approximately five times the load we expect in the SAT under 1$g$. 
For the tension test (configuration \#1), the truss was suspended from a hoist attached to the plate mounted on the 1~K ring of the truss and weight totalling 4.2~kN was suspended from the plate attached to 4~K truss ring.
In both cases care was taken to make sure that an even weight distribution was maintained across the truss.
After each test the truss was visually inspected for damage, and none was observed.
After passing all four tests, the truss was installed in the SAT and has since undergone several cryogenic cycles without issue.


\subsection{Thermal Testing}
\label{sec:thermal_testing}

As many CFRP materials currently on the market are made with proprietary material compositions and fabrication processes, it is important to measure the thermal conductivity at cryogenic temperatures.
The thermal isolation provided by the composite material can vary greatly due to the amount and type of binding polymer, the fiber fill density, and directional layup used with the carbon fiber.
CFRP rods from both sources discussed in Section~\ref{sec:cf_tube} - from Clearwater Composites (CW1 and CW2) and from vDijk Pultrusion Products (DPP) - were thermally characterized via direct cryogenic measurements.


\subsubsection{Clearwater Composites}

The CFRP tubes used for the Clearwater Composites thermal conductivity measurements have a 1.68~mm ID and a 0.38~mm wall thickness.
These tubes are custom made and cut to our desired lengths by the manufacturer.
Figure~\ref{fig:DPP_CW_Graphlite} shows the results of testing the CW tubes in two different configurations.
The first configuration, denoted by ``CW tube'' in Figure~\ref{fig:DPP_CW_Graphlite}, is constructed of a set of twelve tubes cut to 24.4~mm lengths which were epoxied into two copper mounts using Stycast 2850~FT.
This assembly was then mounted to the mixing chamber of a dilution refrigerator with both a ruthenium oxide thermometer and a heater mounted to the thermally isolated end.
The conductivity of the tubes was then measured by applying known amounts of power to the thermally isolated end and measuring its temperature change.
This measurement was performed while the end connected to the thermal bath was held at three different mixing chamber temperatures: 100~mK, 250~mK, and 500~mK.
The next experimental configuration more closely resembles the truss described in this paper. 
The assembly consisted of two sets of CFRP tubes mounted between an intermediate temperature ring.
The first set of eight tubes are 62.2~mm long and the next set consists of eight 25.4~mm long tubes.
This assembly was then mounted to the still of a $^3$He adsorption refrigerator.
The thermal conductivity results, denoted by ``CW assembly'' in Figure~\ref{fig:DPP_CW_Graphlite}, were taken at both 270~mK and 360~mK bath temperatures.
The data agrees across multiple bath temperatures and experimental configurations, ensuring a robust understanding of the thermal properties of the material.
The best-fit $\kappa(T)$ model for the CFRP used in the CW tubes is given in Table~\ref{table:thermalConductivityFits}.


\subsubsection{DPP}

To characterize the thermal conductivity of DPP, test samples were prepared in a similar manner via the setup described in Reference~\cite{Howe_2018}.
Solid rods with a diameter of 10~mm and a length of 5~cm were attached via Stycast 2850~FT epoxy to a copper heat sink on one end and to a small copper tab with a heater and thermometer on the opposite end.
The heat sink was provided by liquid $^4$He and liquid $^3$He baths in two complimentary configurations to extend the dynamic range of the measurement.
By applying varying amounts of Joule power to the floating end of the sample, the thermal conductivity of the sample may be determined from the first derivative of the resulting relationship between the floating side temperature and applied power.

\begin{figure}[t]
\centering
\includegraphics[width=\columnwidth]{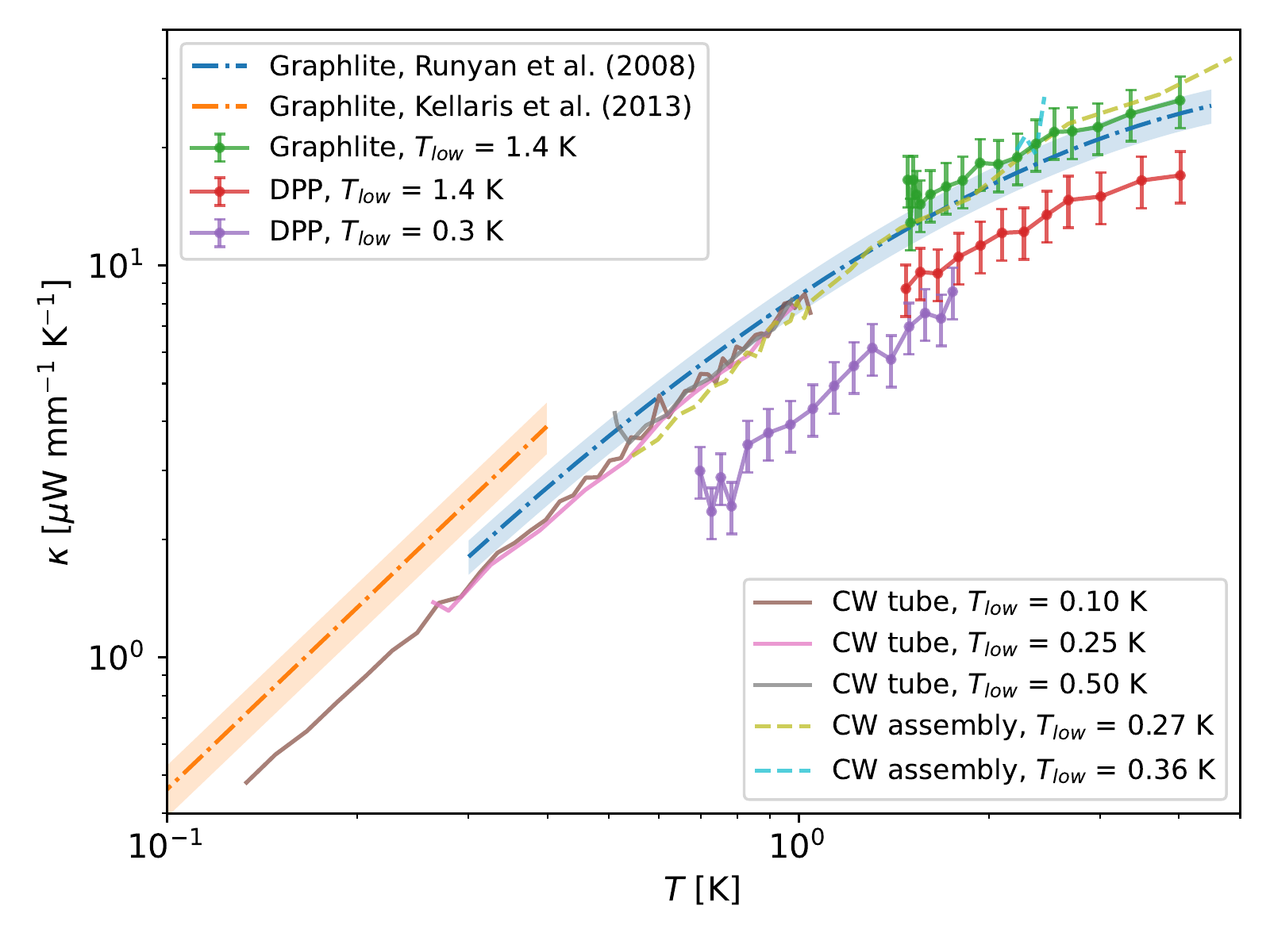}
\caption{
Measured cryogenic thermal conductivities of Clearwater, DPP and Graphlite\textsuperscript{TM}  CFRP rods, along with best-fit curves from published measurements of Graphlite\textsuperscript{TM} \cite{Runyan_2008, Kellaris_2014}.
}
\label{fig:DPP_CW_Graphlite}
\end{figure}

The results of these measurements are shown in Figure~\ref{fig:DPP_CW_Graphlite}.
Due to the relatively small temperature ranges probed in each individual measurement, the data are not sufficiently constraining for power law fits.
Phenomenologically, however, linear fits may be used to estimate thermal loading in practical structures.
The best-fit $\kappa(T)$ models for the DPP tubes are given in Table~\ref{table:thermalConductivityFits}.

As a cross-check, samples of Graphlite\textsuperscript{TM} were also tested side-by-side in the measurement setup.  
Graphlite\textsuperscript{TM} is a similar CFRP that has been used in cryogenic support structures, and its thermal conductivity has been previously measured by other groups \cite{Runyan_2008, Kellaris_2014}.
Our measurement agrees with previous measurements to within uncertainties, giving additional confidence in the DPP measurement results. %
The best-fit $\kappa(T)$ model for Graphlite is given in Table~\ref{table:thermalConductivityFits}.

We ascribe a global 15\% uncertainty to these measurements, resulting from a combination of residual temperature dependence of the heater, the precision of the physical dimensions of the sample, and parasitic thermal paths through wiring and radiation.


\begin{table}
\centering
\footnotesize

\renewcommand\arraystretch{1.3}

\begin{tabular}{ccc}
\hline
CFRP type   & $\kappa(T)$ [\,$\mu \mathrm{W}\, \mathrm{mm}^{-1} \mathrm{K}^{-1}$\,] & Temperature range \\
\hline
CW          & $7.5\,T^{1.2715}$     & $0.1 \, \textrm{K} \leq T  \leq 5.0 \, \textrm{K}$ \\
DPP         & $5.43\,T - 1.27$      & $0.7 \, \textrm{K} \leq T  \leq 1.8 \, \textrm{K}$ \\
DPP         & $3.38\,T + 4.53$      & $1.5 \, \textrm{K} \leq T  \leq 4.0 \, \textrm{K}$ \\
Graphlite   & $4.95\,T + 7.84$      & $1.5 \, \textrm{K} \leq T  \leq 4.0 \, \textrm{K}$ \\
\hline
\end{tabular}

\caption{Best-fit thermal conductivity models.  These models are plotted along with the measurements in Figure~\ref{fig:DPP_CW_Graphlite}.}

\label{table:thermalConductivityFits}
\end{table}


\subsubsection{Loading Prediction}

Given the above measurements for the thermal conductivity of the CW tubes, we can estimate the amount of conductive loading we expect to see from the struts in our truss.
While the CW tubes we used for our thermal conductivity measurements are not identical to the CW we ultimately used in our struts, we assume here that they are sufficiently similar for the purposes of a rough loading estimate.
We estimated the total conductive loading on the 1~K stage of the instrument as
\begin{equation}
\Dot{Q}_{\mathrm{struts}} = N_{\mathrm{struts}} \left( \frac{A_{\mathrm{strut}}}{l_{\mathrm{strut}}} \right) \int_{T_1}^{T_2} \kappa_{CW}(T) \, dT.
\label{eq:thermal_prediction}
\end{equation}
Here, $N_{\textrm{struts}}$ = 24, $A_{\textrm{strut}}$ = 11.8 $\textrm{mm}^2$, $l_{\textrm{strut}}$ = 31.0~mm, $T_1$ = 1~K, $T_2$ = 4~K, and we find $\Dot{Q}_{\textrm{struts}}$ = 0.67~mW.
Given that our BlueFors SD400 DR has a measured still stage cooling power of 30~mW at 1.2~K, this load is well within our thermal budget.


\section{Conclusion}
\label{sec:conclusion}

This study shows that our aluminum end cap design with the flow channels inside the socket works well for this application.
The strength performance is repeatable from strut to strut.
We did not observe any fracturing when the strut samples were thermally shocked with LN2, and the measured strength of the thermally shocked strut samples was the same as the strength of the struts that were not shocked.
The carbon fiber tube variant used in the strut does not seem to affect the overall strength because the limiting elements are the screws and the tab in the end cap, not the end cap socket, the epoxy adhesive, or the carbon fiber tube.
We ultimately chose to use the CW2 tubes for the SAT because they are more robust to handling.
The pull-test measurements indicate the struts will start to yield when the load is approximately 2.0~kN, which means the measured FOS is 5.4 in the worst case for normal operation and 1.4 assuming an unexpected 10$g$ shock during site transport.
Our measurements of the thermal conductivity of the CW2 tube variant suggest that the heat load on the 1~K stage of the instrument through the struts in the truss should be approximately 2\% of the cooling power of the 1~K stage of the DR, which meets our requirement.
The assembled truss has worked as expected both mechanically and thermally in laboratory validation testing of the SAT.


\section{Discussion}
\label{sec:discussion}

There are two discrepancies between the expected and the measured strut performance that are worth noting.
We include possible explanations here.

First, the epoxy sample (see Figure~\ref{fig:epoxy_test}) measurements show that the epoxy joint starts to fail when the tensile load is approximately 3.0~kN, and the epoxy fully fails near 5.0~kN.
Calculations using lap shear values from the epoxy manufacturer show the epoxy joint should be able to hold approximately 6.9~kN at 300~K before cohesively failing, so our measured strength is somewhat less than predicted, even though it greatly exceeds the performance requirement.
The discrepancy (5.0~kN versus 6.9~kN) could come from the bond line thickness.
Our bond line thickness is greater than the bond line thickness used in standard lap shear measurements.
A greater bond line thickness should weaken the joint, which is what we observed.

Second, the M4 stainless steel screws that fasten the struts to the truss rings yielded when the axial load was approximately 3.0~kN.
We calculated that they should yield at 1.8~kN, so our measured result is somewhat better than the prediction.
One possible explanation for this discrepancy is the screws do not fail in shear, as described by distortion-energy theory, but more in tension or a combination of shear and then tension (see failure points in Figure~\ref{fig:screw_test}).
In tension, an M4 screw should yield at 3.1~kN.

A common additional source of heating in CMB instruments comes from low-frequency vibrations that couple to the low-temperature stages of the instrument via the supports.
However, because the nature of this coupling depends strongly on the overall cryostat design and vibrational environment of the instrument, we have not considered it here.
Instead, an analysis of heating from vibrations in the SAT will be described in a future publication (Galitzki \textit{et al}.\ 2022).

The framework we have outlined for the design, fabrication, and assembly of this strut is general.
We have already scaled this design to CFRP struts with OD = 3~mm and ID = 2~mm  for other trusses in the SAT and seen proportional results.
While the nuances of the geometry are different, the procedure for designing the flow channels, preparing the end caps and CFRP tubes, and assembling the struts is the same.
The design we have presented can therefore serve as the foundation for the design of the low temperature supports of future CMB instruments.
Other fields that are also increasing the scale of their low temperature technology, such as quantum computing, might look to this design for future instruments.


\section{Acknowledgements}
\label{sec:acknowledgements}

This work was supported by the Simons Foundation, the University of Pennsylvania, and the University of Virginia.
We would like to thank Natasha Smith at the University of Virginia for letting us use the Instron to test our strut samples.
We would like to thank Jeff Engbrecht at Clearwater Composites for providing useful information about the CW1 and CW2 carbon fiber tubes.
Zhilei Xu is supported by the Gordon and Betty Moore Foundation through grant GBMF5215 to the Massachusetts Institute of Technology.


\bibliography{references}


\appendix


\section{Strut Strength Specification}
\label{sec:strut_strength_specification}

In this appendix we give the full results of the truss FEA (see Section~\ref{sec:truss_design}).
The expected axial and radial force on each strut for each of the five configurations is listed in Table~\ref{table:truss_loads}.
Note that a positive axial force corresponds to a strut in tension, and a negative axial force corresponds to a strut in compression.
All radial forces are positive by definition.


\begin{table*}[t]
\footnotesize
\centering
\renewcommand\arraystretch{1.2}
\begin{tabular}{c|cc|cc|cc|cc|cc}
\multicolumn{1}{c}{ } & \multicolumn{2}{c}{cfg.\ \#1 (el = -90~deg)} & \multicolumn{2}{c}{cfg.\ \#2 (el = -45~deg)} & \multicolumn{2}{c}{cfg.\ \#3 (el = 0~deg)} & \multicolumn{2}{c}{cfg.\ \#4 (el = 45~deg)} & \multicolumn{2}{c}{cfg.\ \#5 (el = 90~deg)}\\
\hline
strut   & axial & radial    & axial & radial    & axial & radial    & axial & radial    & axial & radial    \\
\hline
a       & 70.4  & 1.61      & 137. & 4.00       & 123.  & 4.15      & 37.3  & 1.88      & -695. & 15.0      \\      
b       & 108.  & 3.60      & 260. & 7.27       & 260.  & 6.25      & 107.  & 2.41      & -1070 & 35.9      \\      
c       & 117.  & 3.57      & 114. & 7.62       & 44.3  & 7.35      & -51.1 & 2.63      & -1150 & 34.8      \\
d       & 98.1  & 2.45      & 259. & 3.58       & 269.  & 3.55      & 121.  & 2.72      & -969. & 23.7      \\
e       & 144.  & 2.83      & 5.61 & 5.98       & -136. & 5.72      & -198. & 2.47      & -1420 & 28.8      \\
f       & 138.  & 3.24      & 350. & 3.04       & 358.  & 3.32      & 156.  & 3.10      & -1370 & 32.7      \\      
g       & 99.8  & 2.30      & -74.7  & 5.14     & -205. & 5.49      & -216. & 3.17      & -985. & 22.0      \\      
h       & 109.  & 3.50      & 230. & 1.66       & 217.  & 4.97      & 76.5  & 5.55      & -1070 & 34.9      \\      
i       & 109.  & 3.46      & -162.  & 2.57     & -337. & 3.70      & -316. & 4.19      & -1070 & 33.6      \\      
j       & 97.6  & 2.27      & 88.5 & 3.30       & 27.6  & 6.79      & -49.5 & 6.27      & -962. & 22.1      \\      
k       & 136.  & 2.97      & -161.  & 2.21     & -363. & 5.18      & -352. & 5.60      & -1340 & 30.2      \\      
l       & 137.  & 2.82      & -8.67  & 2.62     & -150. & 6.94      & -202. & 6.73      & -1360 & 29.7      \\      
m       & 95.6  & 2.20      & -55.4  & 2.20     & -174. & 5.08      & -190. & 5.43      & -942. & 21.7      \\      
n       & 109.  & 3.41      & -108.  & 2.43     & -261. & 6.11      & -262. & 6.41      & -1080 & 33.3      \\      
o       & 107.  & 3.44      & 51.3 & 2.68       & -34.1 & 7.23      & -100. & 7.47      & -1050 & 34.6      \\
p       & 98.0  & 2.25      & -121.  & 2.80     & -269. & 3.55      & -259. & 3.58      & -966. & 21.9      \\      
q       & 137.  & 2.84      & 201. & 2.20       & 147.  & 5.83      & 6.95  & 5.93      & -1360 & 29.7      \\      
r       & 137.  & 2.82      & -155.  & 3.42     & -356. & 3.60      & -349. & 3.08      & -1360 & 29.1      \\      
s       & 100.  & 2.27      & 223. & 3.01       & 214.  & 6.17      & 80.5  & 5.26      & -988. & 22.3      \\      
t       & 108.  & 3.54      & -76.7  & 5.79     & -217. & 5.00      & -229. & 1.89      & -1070 & 34.5      \\      
u       & 114.  & 3.47      & 327. & 4.32       & 348.  & 4.10      & 165.  & 2.90      & -1130 & 35.5      \\      
v       & 98.4  & 2.39      & 50.2 & 6.50       & -27.5 & 7.02      & -89.0 & 3.26      & -972. & 23.7      \\      
w       & 146.  & 3.04      & 367. & 5.80       & 372.  & 5.31      & 160.  & 2.24      & -1450 & 31.8      \\      
x       & 138.  & 3.36      & 203. & 7.25       & 150.  & 6.82      & 8.65  & 2.58      & -1360 & 27.4      \\ 
\hline
\end{tabular}
\caption{
Expected radial and axial strut forces for five limiting-case truss configurations computed with FEA.
The units are newtons.
Note that configuration~\#5 was simulated with 10 times the true load to simulate an unexpected drop of the instrument during site transport.
The strut letter here corresponds to strut letter in Figure~\ref{fig:TrussSimOrientation}.
}
\label{table:truss_loads}
\end{table*}


\begin{figure}[t]
\centering
\includegraphics[width=\columnwidth]{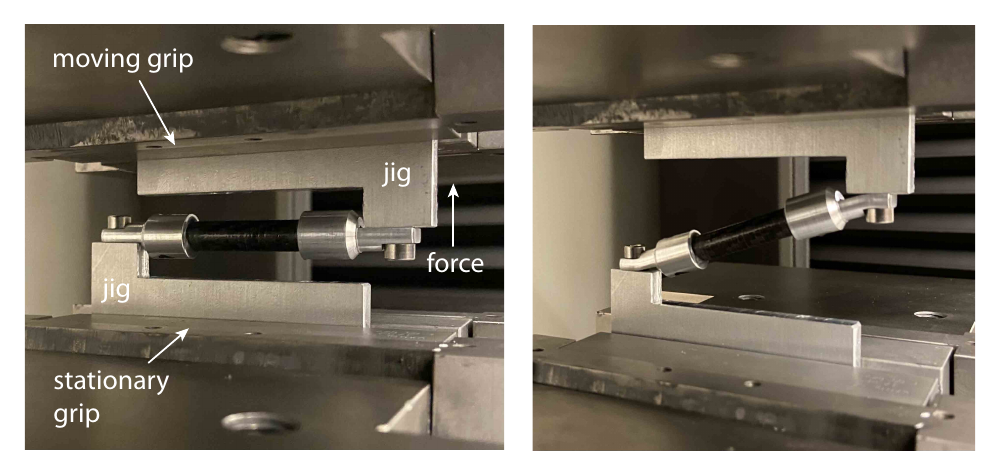}
\caption{
Left:
Photo of the shear test setup before the force was applied.
Right:
Photo of the strut when the measurement was stopped.
For scale, the dimensions of the strut are given in Figure~\ref{fig:strut_drawing}.
}
\label{fig:shear_test_setup}
\end{figure}


\begin{figure}[t]
\centering
\includegraphics[width=0.4\textwidth]{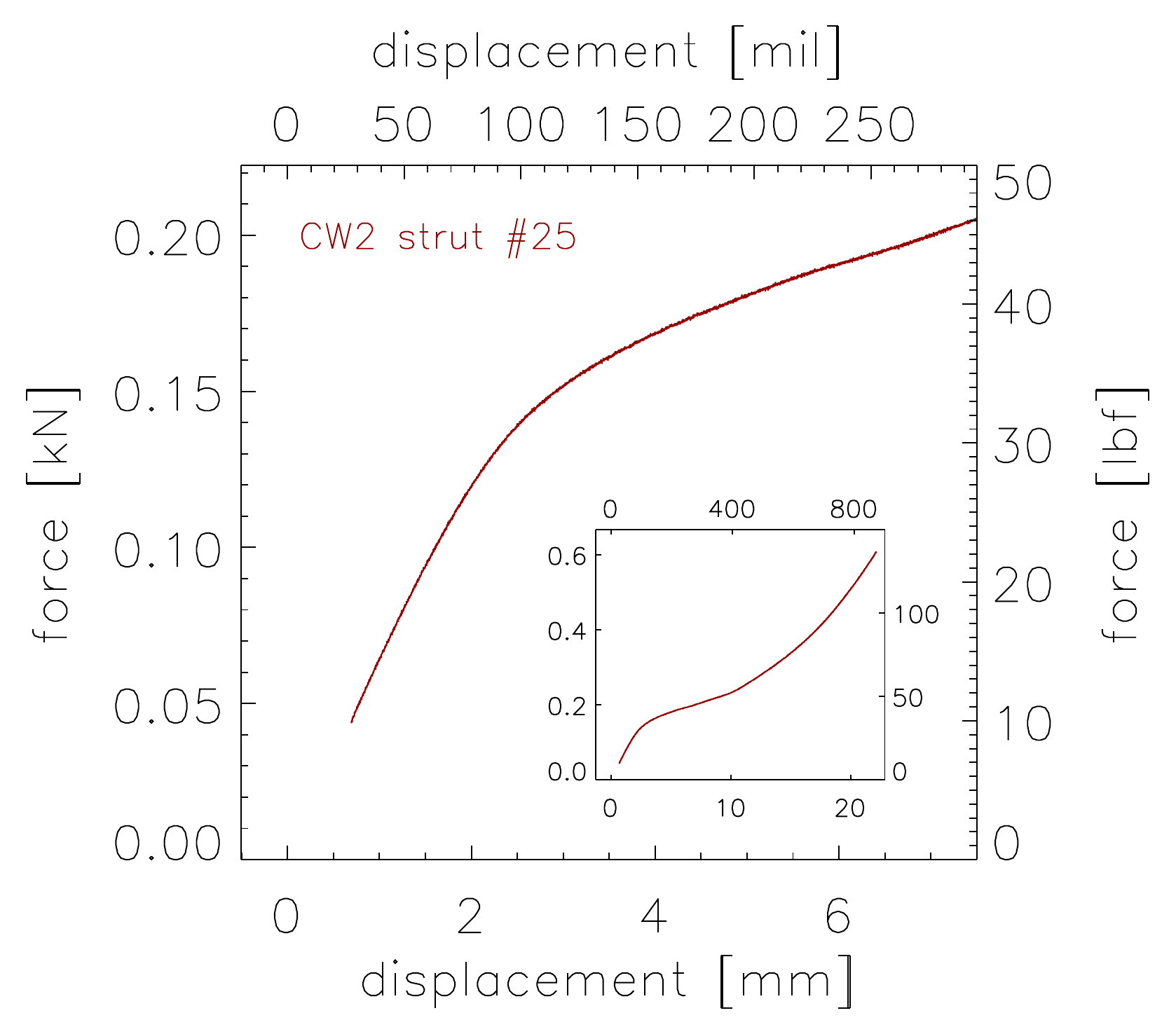}
\caption{
Results from the strut shear test.
A force versus displacement measurement for one CW2 strut is plotted.
The most relevant part of the data appears in the main plot and the full data set is plotted in the inset for completeness.
The aluminum end cap starts to yield when the displacement is approximately 2~mm.
For the SO truss, the worst-case deflection is expected to be 1.5~mm (see Section~\ref{sec:truss_testing}).
}
\label{fig:shear_test}
\end{figure}


\begin{figure}
\centering
\includegraphics[width=0.4\textwidth]{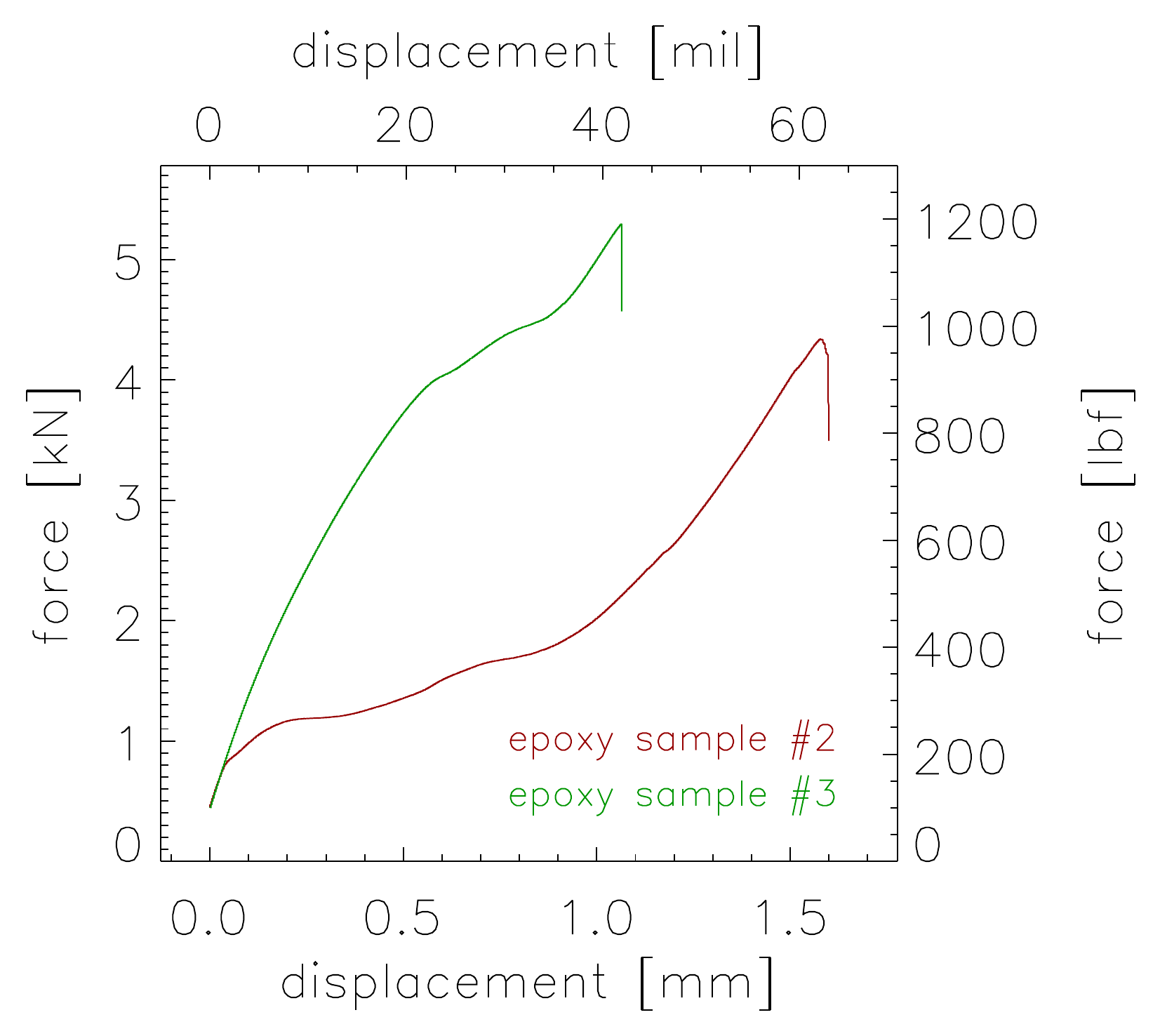}
\caption{
Force versus displacement measurement for two CW2 struts tested in compression.
We suspect the general shape of the two curves is produced by the way the carbon fiber tube moved through the broken epoxy in the end cap socket, so this part of the measurement is likely irrelevant.
The sharp feature at the top of the two curves is produced when the carbon fiber tube breaks.
There is no compression-strength prediction for the CW2 tubes, but this result is similar to the tensile-strength prediction in Table~\ref{table:cf_tube_info}. 
}
\label{fig:compression_test}
\end{figure}


\section{Ancillary Tests}
\label{sec:ancillary_tests}

In addition to the tests described in Section~\ref{sec:strut_testing}, we ran some additional tests that were less conclusive but the results are worth reporting.
We include the results from these additional tests in this appendix.

The fifth CW2 strut (strut \#25) was tested in shear.
For this test, we made a separate mounting jig for the Instron, which is shown in Figure~\ref{fig:shear_test_setup}.
This jig initially stressed the strut in shear and then in tension as the displacement grew.
The measurement is plotted in Figure~\ref{fig:shear_test}, and it shows that the end caps remain in the elastic limit for approximately the first 2~mm of displacement.
This result suggests that the differential thermal contraction effect described in Section~\ref{sec:truss_testing} will not cause the struts to fail.
We stopped the test when the displacement grew to approximately 20~mm.
None of the strut elements failed, though the two end caps did plastically deform. 

After the four epoxy samples were pull tested to failure, we decided to use two of them (samples \#2 and \#3) to also test the CW2 tubes in compression.
Figure~\ref{fig:epoxy_test} shows a photo of one epoxy sample after the epoxy joint was broken and the carbon fiber tube was pulled from the end cap.
Neither the carbon fiber tube nor the end cap were damaged, so for the compression test, we reinserted the carbon fiber tube back into the end cap and then installed the compression specimen in the Instron grips.
The data from the two compression measurements are plotted in Figure~\ref{fig:compression_test}.
In compression, the CW2 tube in sample \#2 broke at approximately 5.3~kN, and the CW2 tube in sample \#3 broke at approximately 4.4~kN.
There was no prediction for the strength of the CW2 tubes in compression, but this result suggests that the CW2 tubes are strong enough for this application given the FEA results in Table~\ref{table:truss_loads}.


\end{document}